\documentclass[journal]{paperlayouttran}

\usepackage{cite}

  \usepackage{color}
\usepackage[normalem]{ulem}

\ifCLASSINFOpdf
  \usepackage[pdftex]{graphicx}
  \graphicspath{{images/}}
\else
  \usepackage[dvips]{graphicx}
  \graphicspath{{images/}}
  
\fi

\ifCLASSOPTIONcompsoc
 \usepackage[caption=false,font=normalsize,labelfont=sf,textfont=sf]{subfig}
\else
 \usepackage[caption=false,font=footnotesize]{subfig}
\fi

\usepackage{siunitx}
\usepackage{dsfont}
\usepackage{amssymb}
\usepackage{amsmath}
\usepackage{multirow}

\begin{document}
\title{Automatic Mapping of Atrial Fiber Orientations for Patient-Specific Modeling of Cardiac Electromechanics using Image-Registration}

\author{Julia~M.~Hoermann,
	Martin~R.~Pfaller,
	Linda~Avena,
	Crist\'obal~Bertoglio$^\dag$,
        and~Wolfgang~A.~Wall$^\dag$
\thanks{Julia M. Hoermann, Martin R. Pfaller and Wolfgang A. Wall are with the Institute
of Computational Mechanics, Technical University of Munich, Boltzmannstra\ss{}e 15, 85748 Garching bei M\"unchen,
Germany. e-mail: hoermann@lnm.mw.tum.de}% <-this % stops a space
\thanks{Crist\'obal~Bertoglio is with the Bernoulli Institute, University of Groningen, Nijenborgh 9, 9747 AG Groningen, Netherlands.}
\thanks{Linda Avena is with the Department of Electrophysiology, German Heart Center Munich, Technical University of Munich, Lazarettstr. 36, 80636 M\"unchen, Germany.}% <-this % stops a space
\thanks{$^\dag$ Joint last authors}
}

\markboth{International Journal for Numerical Methods in Biomedical Engineering} %,~Vol.~14, No.~8, August~2015}%
{}

\maketitle

\begin{abstract}
Knowledge of appropriate local fiber architecture is necessary to simulate patient-specific electromechanics in the human heart. However, it is not yet possible to reliably measure in-vivo fiber directions especially in human atria. Thus, we present a method which defines the fiber architecture in arbitrarily shaped atria using image registration and reorientation methods based on atlas atria with fibers predefined from detailed histological observations. Thereby, it is possible to generate detailed fiber families in every new patient-specific geometry in an automated, time-efficient process.
We demonstrate the good performance of the image registration and fiber definition on ten differently shaped human atria. Additionally, we show that characteristics of the electrophysiological activation pattern  which appear in the atlas atria also appear in the patients' atria.  We arrive to analogous conclusions for coupled electro-mechano-hemodynamical computations. % keep it simple ...
\end{abstract}

\begin{paperlayoutkeywords}
atrial fiber orientation; patient-specific modeling; cardiac electromechanics.
\end{paperlayoutkeywords}

\paperlayoutpeerreviewmaketitle

\section{Introduction}
\paperlayoutPARstart{P}{atient-specific} mathematical and computational models can contribute to understand the (patho-) physiological function of the heart. 
 These models require not only an accurate geometrical representation of the heart, usually obtained from computed tomography (CT)  or from magnetic resonance imaging (MRI), but also a description of the fiber directions. The term fiber is referred to myofiber bundles, which are similarly oriented myocytes running along a certain direction denoted as fiber direction. For a correct representation of the electrophysiological behavior knowledge about the fiber direction is necessary, since the  electrical signal travels faster in fiber direction than perpendicular to it \cite{clerc_directional_1976} and this anisotropy influences the electrical activation \cite{zhao_imagebased_2012}. But a correct fiber representation is obviously also important from the mechanical point of view, as e.g. studies of the left ventricle show that different fiber architecture lead to different results in the mechanical contraction due to active and passive anisotropy \cite{eriksson_influence_2013,nikou_sensitivity_2016}.

The fiber architecture of the atria differs from the one of the ventricles. While in the ventricles the fibers are aligned in an almost regular way \cite{streeter_fiber_1969,ennis_myofiber_2008}, the fibers in the atria are aligned in individual bundles which run in different directions through the left and right atrial wall \cite{ho_atrial_2002}. Due to individual fiber bundles running in different directions it is not straight-forward to use rule-based approaches, 
as it can be done in the ventricles \cite{bayer_novel_2012,wong_generating_2014}. Besides rule-based methods, another promising approach to define the fiber direction in ventricles is to use diffusion tensor MRI (DTMRI), which is capable to measure non-invasively the fiber architecture of the left ventricular myocardium \cite{hsu_magnetic_1998}. This technology is often used ex-vivo \cite{helm_ex_2005,peyrat_computational_2007,nagler_personalization_2013}, since in-vivo measurements are challenging \cite{sosnovik_diffusion_2009,stoeck_second-order_2016} and only few slices can be acquired. Furthermore, it requires sophisticated reconstructions of the fibers for the whole ventricles \cite{toussaint_vivo_2013,nagler_maximum_2017}.  
However, until now it is not possible to obtain in-vivo fiber directions in the atria since their thickness is smaller than current DTMRI voxel size. Precisely, the atrial wall is around \SI{2}{\milli\metre} thick \cite{beinart_left_2011}, while in comparison the left ventricle is around \SI{8}{\milli\metre} thick \cite{kawel_normal_2012}. Only recently, ex-vivo fiber orientation in eight different atria could be analyzed with submillimeter DTMRI \cite{pashakhanloo_myofiber_2016}, which could be used as additional information for fiber definitions in the future.

Until now, only few methods have been proposed to create fiber directions in patient-specific atria \cite{sachse_model_1999,hermosillo_semiautomatic_2008,krueger_modeling_2011}. The approach in \cite{krueger_modeling_2011} is a first step towards realistic atrial models. The importance of fiber orientations for patient-specific simulations is demonstrated with different geometrical models, to which the semi-automatic method is applied and additional electrophysiological simulations are performed. The method uses voxel based atrial images and manually defined seed points to specify different fiber bundles using marching level sets methods. It is shown that this method works very robust and the results correspond well with reported literature. However, the semi-automatic approach is strongly depending on user input of the 22 seed points, which have to be set in an accurate way. Variation of the seed points can lead to different results. Strong shape variations, which are common in human atria, are difficult to 
handle with this algorithm, since it depends on shortest paths between seed and auxiliary points and subdivisions at fixed relations. For example to incorporate absent or additional pulmonary vein orifices an adaption of the algorithm is necessary.  Additionally, these models are limited to the information about anatomical fiber orientations known at the time of the development of the methods. 
New information can be only incorporated into the model by changing the methodology, which in turn would lead to more user interaction by defining additional seed points.

To the authors'  best knowledge until now image registration techniques are not yet used for fiber estimation in the atria. We propose a method to map atlas atria to a patient's atria using image registration techniques. Using the computed deformation map, the fiber directions are reoriented and then transferred to the patient's atria. The initial user input of seed point is reduced in comparison to rule-based models and is only needed for a first general alignment of the geometries. The influence of user variations is therefore greatly decreased. Additionally, the benefit of using registration methods is that the accuracy of the fiber orientation can be easily improved by adapting only the atlas model. More complex data and details have to be included only in the atlas model without changing the registration algorithm.
Additionally, geometry variations as the number of pulmonary orifices, which is a challenge for the rule based models, can be handled by the usage of different atlas atria. Another benefit is the possibility of using ex-vivo measured DTMRI atria with fiber data as atlas atria, thus, improving the accuracy of the model.
In \cite{vadakkumpadan_imagebased_2012} a method has been proposed to register an atlas ventricle to the patient's ventricle using large deformation diffeomorphic metric mapping. The goal of this paper is to propose an image registration and reorientation method for atrial geometries and fibers and to demonstrate its performance solving an electromechanical problem of patient-specific atria using our fiber definition.

The reminder of the paper is organized as follows. In Section \ref{sec_methods} we characterize the method to define the fibers of the atlas atria. Then we describe the process of image registration, fiber reorientation and fiber lifting to generate the fibers in the patient's atria.
Additionally, we shortly describe the electro-mechano-hemodynamical model, with which we simulate the functions of the atria with defined and mapped fibers. In Section \ref{sec_results} we describe the results of the registration and mapping procedure in comparison to the defined fibers. Additionally, we compare, analyze and discuss the results of the electromechanical simulation with mapped fibers in all atria.

\section{Methods}
\label{sec_methods}

\begin{figure*}[!t]
\centering
\includegraphics[width=0.8\textwidth]{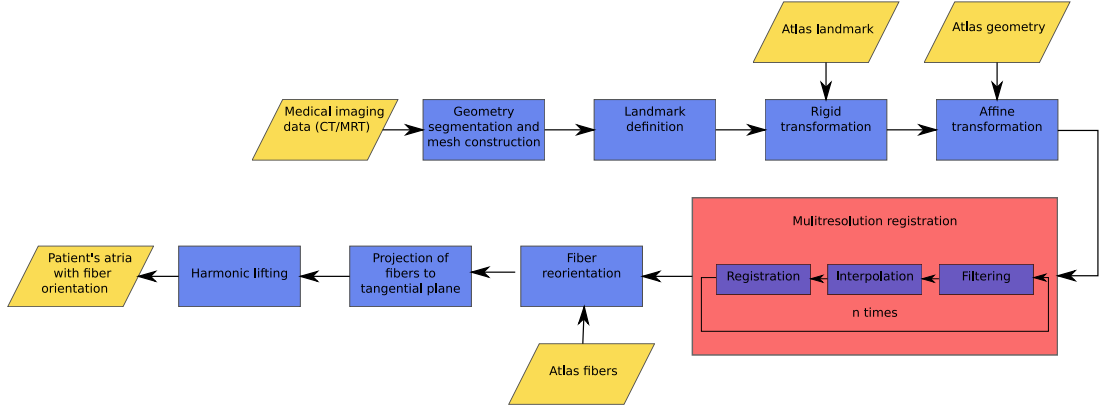}
\caption{Processing pipeline for the registration and fiber estimation for the patients' atria}
\label{fig_flowchart}
\end{figure*}

Several steps are necessary for the estimation of the fiber architecture in patient's atria. Firstly, an atlas atria with a physiologically detailed fiber architecture has to be created. This is done once at the beginning and the atlas is then used for fiber estimation for all atrial geometries. For each patient the following procedure is performed, see Figure \ref{fig_flowchart}. From medical imaging data, a geometry is created, which is registered to the atlas. Then, the deformed atlas with reoriented fibers is used to generate the fibers of the patient. Finally, the fibers are realigned so that they are tangential to the surface and at last a harmonic lifting is performed.

At the end of this chapter we will also briefly describe how computations using an electro-mechano-hemodynamical model are performed with the results of the fiber generation as geometrical input.

\subsection{Geometry Creation}
\label{sec_geometry-creation}
To construct the geometry of a patient's atria we use segmentation, design modification and meshing tools. 
First, a surface representation is generated using the software Mimics (Materialise, Leuven, Belgium). For this, the lumen of both atria is segmented manually so that the endocardial surfaces are obtained. To generate the epicardial surface we extrude the endocardial surface by \SI{2}{\milli\metre}, which corresponds to an average thickness of the atrial wall \cite{beinart_left_2011}. Additionally, we add an interatrial muscular bridge between the right and the left atrium, the Bachmann bundle, to allow a physiological propagation of the electrical signal. Finally, we create a 3D volume mesh with tetrahedral elements with a maximal element size of \SI{0.9}{\milli\metre} using Gmsh \cite{geuzaine_gmsh_2009}.  This leads to about two to three elements through the atrial wall. Note that this does not pose accuracy issues in the computations, since we use higher order spatial discretizations.

\subsection{Fiber Definition for Atlas Atria}
\label{fiberatria}
For the atlas atria we define the fiber orientation using reported detailed knowledge of atrial fiber structure \cite{ho_atrial_2002}. The geometry of the atlas are the atria of a 71-old individual without known cardiac pathological findings. The model was obtained from a cardiac CT image with a resolution of \SI{0.4 x 0.4 x 0.8}{\milli\metre}.
The atlas geometry is then created identically as any other patients' atria and as has just been described in Section \ref{sec_geometry-creation}.

\begin{figure}[!t]
\centering
\subfloat[Posterior view]{\includegraphics[width=0.24\textwidth]{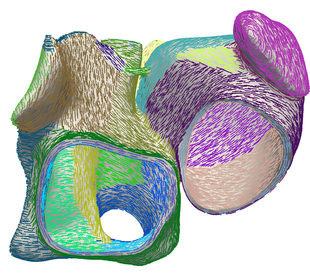}%
\label{fig_first_case}}
\hfil
\subfloat[Anterior view]{\includegraphics[width=0.24\textwidth]{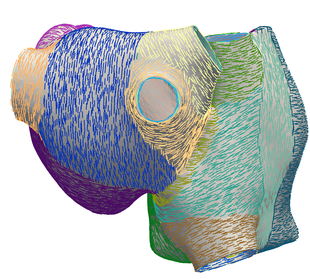}%
\label{fig_second_case}}
\caption{Fiber direction of atlas atria. The colors represent the regions with different fiber directions.}
\label{fig_fiber-atlas}
\end{figure}

To define the fibers in the atlas atria we divide the epicardial and endocardial wall of the left and right atrium into regions with a constant fiber direction (see colored regions in Figure \ref{fig_fiber-atlas}). In doing so, we manually set %considered 
the main fiber bundles crista terminalis, pectinate muscles, circumferential vestibule fibers and the Bachmann bundle. Additionally, also other prominent bundles and fiber alignment in the right and left atrium are defined. In the right atrium endocardial and epicardial fiber directions coincide, i.e. the fiber bundle direction is constant throughout the whole thickness of the wall. In contrary, different fiber bundles throughout the thickness of the wall run in the left atrium at the posterior wall. To model this we assign different fiber directions on the epicardial and endocardial surface. After defining a fiber direction on each node on the surface, we interpolate the fibers into the volume using harmonic lifting techniques \cite{nagler_personalization_2013}. The atlas atria with fiber directions and the different regions can be seen in Figure \ref{fig_fiber-atlas}.

\subsection{Atlas Geometry Registration}
The deformation of the atlas atria to the shape of the patients' atria is done in two major steps. First, an affine transformation is calculated and second the actual elastic registration process is computed. The registration process is performed in MATLAB (Release 2017a, The MathWorks, Inc., Natick, Massachusetts, United States)
\subsubsection{Affine Transformation}
\begin{figure*}[!t]
\centering
\subfloat[]{\includegraphics[trim=0 0 0 0,clip,width=0.32\textwidth]{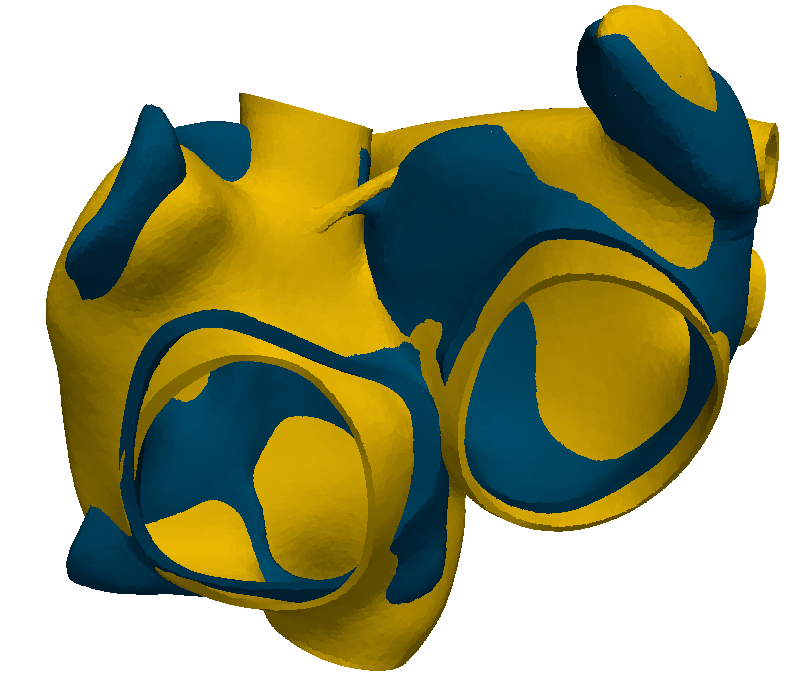}%
\label{fig_global_registration}}
\hfil
\subfloat[]{\includegraphics[trim=0 0 0 0,clip,width=0.32\textwidth]{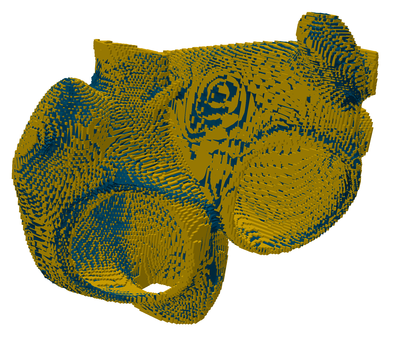}%
\label{fig_atlas_binary_registration}}
\hfil
\subfloat[]{\includegraphics[trim=0 0 0 0,clip,width=0.32\textwidth]{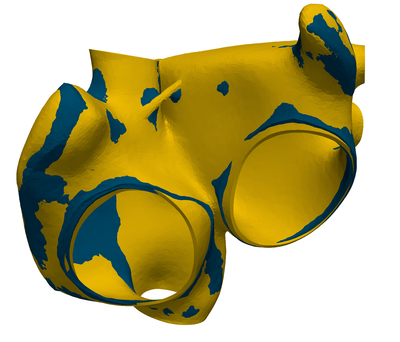}%
\label{fig_atlas_mesh_registration}}
\caption{Atlas atria (blue) and patient's atria (yellow) at different steps of the registration process: (a) meshed geometry after affine transformation,  (b) binary images after registration and (c) the meshed geometry after the registration and interpolation to the tetrahedral nodes.}
\label{fig_binary}
\end{figure*}

For the affine transformation, 13 landmarks on the atlas and the patients' geometry are defined around the orifices of pulmonary veins and around the valvular orifices to the ventricles and the Bachmann bridge (Figure \ref{fig_landmark}).
First, we compute a rigid motion that aligns optimally in a least square sense the two sets of landmark points using Singular Value Decomposition. Note that the landmarks are only used for the rigid transformation, which is needed to generally align the two geometries. Furthermore, three landmarks are sufficient for a unique rigid transformation. Second, we perform an Iterative Closest Point affine registration for the sets of all mesh points of the geometries. Figure \ref{fig_global_registration} shows the atria of Patient 1 and the atlas atria after calculating and applying the rigid and the affine transformation to the atlas atria. 
\subsubsection{Registration}
\begin{figure*}[!t]
\centering
\subfloat[]{\includegraphics[trim=0 0 0 0,clip,width=0.32\textwidth]{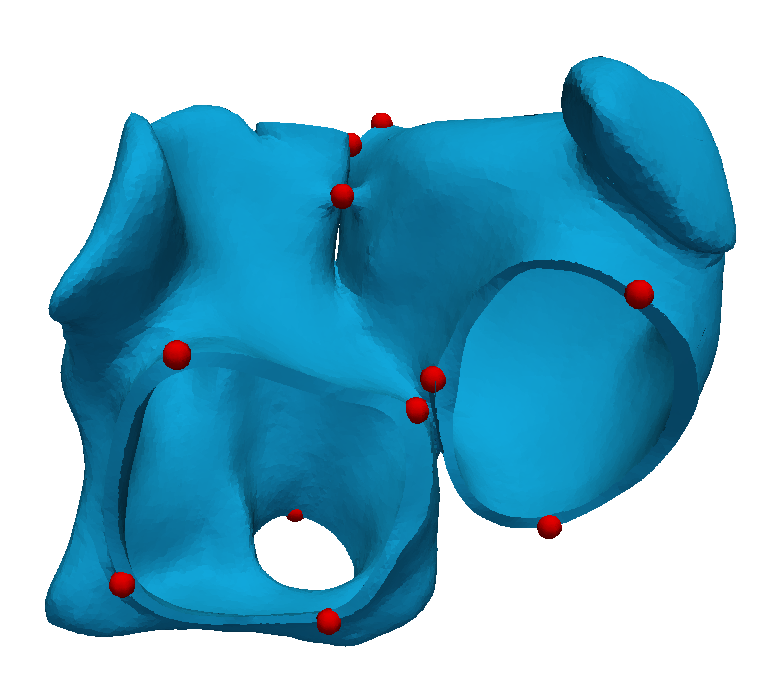}%
\label{fig_landmark}}
\hfil
\subfloat[]{\includegraphics[trim=0 0 0 0,clip,width=0.32\textwidth]{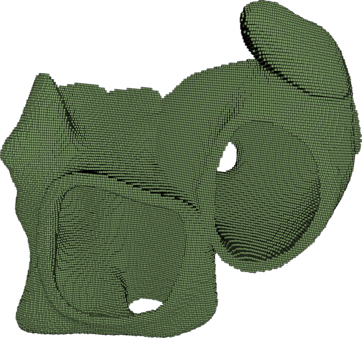}%
\label{fig_atlas_binary}}
\caption{(a) The landmarks defined on the atlas atria. (b) The binary image of the atlas.}
\label{fig_landmark_reg}
\end{figure*}

For the registration we use an algorithm which we already successfully used for material modeling and parameter identification of biological materials \cite{bel-brunon_numerical_2014}. To match the atlas geometry to the patients geometry we first create a 3D binary voxel representation of the atria with a voxel size of \SI{0.6}{\milli\metre} (see Figure \ref{fig_atlas_binary}). The walls in the binary image are slightly thicker than the original walls in the mesh. Nevertheless, this does not yield a problem, since the interpolation is performed with a weighted nearest neighbor principle. This is done for both the atlas atria and the patient's atria.

We denote the image of the patient's atria by $I_p:\Omega\rightarrow \{0,1\}$ and the image of the atlas atria by $I_a:\Omega\rightarrow \{0,1\}$, where $\Omega\subset \mathds{R}^3$ is the 3D image cube. We want to find a transformation $\varphi$ such that the deformed atlas $I_a$ is similar to the patient's atria $I_p$. We use the standard distance measure sum of squared differences (SSD), which 
is given by 
\begin{align}
 \mathcal{D}(\varphi):=\frac{1}{2} \int_\Omega (I_a(\varphi(x))-I_p(x))^2 \, \mathrm{d}\mathbf{x} \text{,}
\end{align}
with  $\varphi(x)=x+u(x)$ and $ u(x):\mathds{R}^3\rightarrow\mathds{R}^3$ a spatially varying displacement field. To overcome the inherent ill-posedness of the image registration problem \cite{fischer_illposed_2008}, an elastic  potential as regularization $\mathcal{S}$ is added \cite{haber_multilevel_2006}. 
Therefore, we formulate the registration problem as: find a displacement field $u^*$ such that 
\begin{align}
  u^* = 
  \arg \min 
   \left[\mathcal{D}(u) + \alpha \mathcal{S}(u)\right]
\end{align}
with the regularization parameter $\alpha>0$. We use a regularization parameter of $\alpha=0.1$ in this work as suggested in \cite{bel-brunon_numerical_2014}. 
To solve the optimization problem we use a Gauss-Newton method \cite{haber_multilevel_2006}. Additionally we use a multiresolution  approach \cite{lester_survey_1999}. From the binary images several levels of coarse grids are consecutively created using convolution with a Gaussian smoothing function. The smoothed image is hereby resampled to an image at a coarser scale with doubled voxelsize. We start the minimization at the coarsest level. The result of the minimization at level $n$ is then linearly interpolated to the grid at level $n-1$ and this result is used as a starting point at the new level. This minimization-interpolation steps are performed at each level until at the finest level, the original binary image, the final transformation $u^*$ is determined. In our case we use $n=4$ levels of resolution.

\subsubsection{Interpolation and Atlas Fiber Reorientation}
The last step in the fiber estimation process is to transfer the fibers of the atlas atria to the patient's atria. To do this we first calculate the transformation of each mesh node from the optimal transformation $u^*$ of the voxels. For each mesh node the nearest voxels are determined. Using the transformation defined in these voxels, the transformation of the mesh node is computed using the inverse distance weighting average of the voxel centers. Additionally, the deformation gradient is calculated at each node using the affine transformation matrix and the Jacobian of the displacement field $u$. To reorient the fiber directions, two strategies are possible, the finite strain strategy and the strategy of principal directions \cite{alexander_spatial_2001}. The finite strain strategy uses the rotational component of the deformation gradient to reorient the fibers. Whereas, the strategy of preservation of principal directions takes also into account the deformation component of the affine transformation. 
Thus, the 
whole deformation gradient is used for the reorientation. In this paper we use the strategy of principal direction. To map the fiber orientation of the deformed atlas atria to the patient's atria we use again an inverse distance weighting of the mesh nodes of the geometries. We map only the fiber orientation to the surface of the patient's atria. Then we project the fiber orientation to the 
tangential plane of the surface nodes and perform afterwards a harmonic lift step to compute the fiber orientation in the interior of the atrial wall \cite{nagler_personalization_2013}. To improve readability, for now on we will use the keyword \textit{mapped} for the fibers, which are generated on the patients' atria through registration and interpolation techniques.

\textcolor{black}{In all our computations, we used n=4 levels of resolution in the registration algorithm.}

\subsection{Electro-mechanico-hemodynamical Computations}

The details of the electrophysiological, mechanical and hemodynamical models are described in previous work \cite{hormann_multiphysics_2017}. For the electrophysiological part, we calculate the monodomain equations with the minimal cell model from \cite{bueno-orovio_minimal_2008}. The parameter set of the cell model is adapted to reproduce atrial activation and a physiological action potential curve for the atrial cell according to  \cite{lenk_initiation_2015}.
The diffusivity is assumed transverse isotropic with a diffusion coefficient of $\sigma_1=\SI[per-mode=symbol]{0.1}{\milli\metre^{2}\per\milli\second}$ in fiber direction and a diffusion coefficient of $\sigma_2=\SI[per-mode=symbol]{0.01}{\milli\metre^{2}\per\milli\second}$ perpendicular to it. To receive physiological propagation we use high-order Hybridizable Discontinuous Galerkin (HDG) methods \cite{hoermann_adaptive_2018}  with a maximal order of five \textcolor{black}{and a stabilization parameter of $\tau=\SI[per-mode=symbol]{1}{\milli\metre\per\milli\second}$}. For time discretization we use a semi-implicit discretization \cite{whiteley_efficient_2006,fernandez_decoupled_2010}, i.e. linear terms implicit and the non-linear parts explicit in time with at time step of \SI{0.1}{\milli\second}. 

The mechanical model is coupled to the electrical model via the electrical signal which triggers the contraction. The model is based on nonlinear elastodynamic equations with a passive and an active part. The passive material is modelled as nearly incompressible Mooney-Rivlin material, \textcolor{black}{where  a volumetric penalization technique for the incompressibility was used. Rayleigh-type damping was also included.}
The active part is described by an active stress component \cite{bestel_biomechanical_2001,chapelle_energypreserving_2012}. The maximal active tension is defined as \SI{100}{\kilo\pascal}. The computation of the mechanical model is performed using tetrahedral quadratic continuous finite elements for space discretization and \textcolor{black}{the generalized-$\alpha$ method for time discretization}. 

For the hemodynamical model we use a three-element Windkessel model  \cite{shi_review_2011}, which is coupled monolithically with the elastic myocardial wall \cite{hirschvogel16}. \textcolor{black}{We used Windkessel models at the orifices to the ventricles. For all atrial models we used for the sake of simplicity the same set of parameters, which are adjusted for the Patient 1 using ventricular ejection fraction captured using cine MRI images.}

\section{Results}
\label{sec_results}
To demonstrate the performance of our method we investigate on one hand the difference between mapped and defined fibers and on the other hand we demonstrate the functionality of our method on ten different patients' atria. We enumerate the atria of the patients by Patient 1 to Patient 10. 

In the following part A, we show a comparison between mapped fibers and defined fibers. For that, we manually define for Patient 1 the fiber orientation, performing the same steps as for the atlas atria described in Section \ref{fiberatria}. Then, we first map the atlas fibers to Patient 1 and second, we map the fibers of Patient 1 to the atlas atria. Afterwards, we compare the fiber orientations, the electrophysiological activation and the mechanical contraction between the two pairs of mapped fibers and manually defined fibers. 

In the second part, the fiber orientation for ten differently shaped atria are generated and the electrophysiological activation pattern and the contractions are computed.

%\textcolor{black}{The runtime of the registration and fiber orientation depends strongly on the size of the atria, since the voxel size was the same for all geometries. The overall registration and fiber definition takes between a few hour to around a day to complete.}

\subsection{Comparison of Defined and Mapped Fibers}

\subsubsection{Fibers}
\label{sec_fibers}
\begin{figure}[]
\centering
\subfloat[]{\includegraphics[trim=0 0 0 0,clip,width=0.22\textwidth]{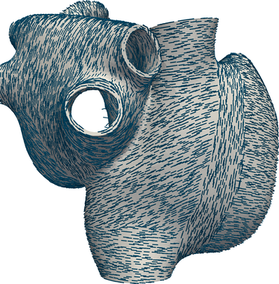}%
\label{fig_fiber_eva_reg}}
\hfil
\subfloat[]{\includegraphics[trim=0 0 0 0,clip,width=0.22\textwidth]{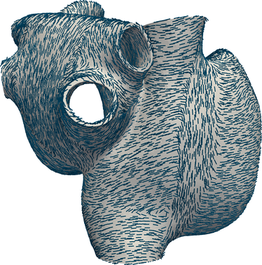}%
\label{fig_fiber_eva_def}}
\hfil
\subfloat[]{
\begin{minipage}{0.5\textwidth}
\centering
                \includegraphics[width=0.44\textwidth]{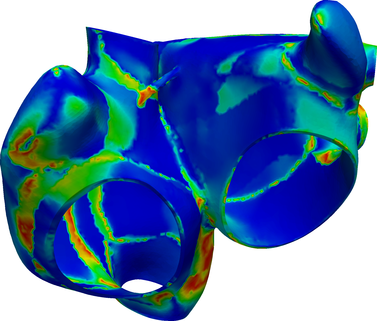}
                \includegraphics[width=0.44\textwidth]{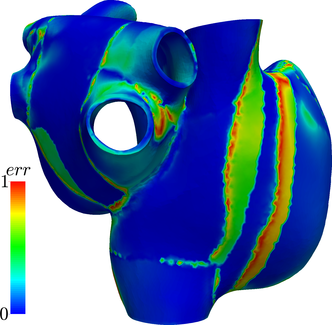}
            \end{minipage}%
\label{fig_eva_diff}}
\caption{Patient 1 fibers.  (a) Defined fiber orientations in Patient 1. (b) Mapped fiber orientation in Patient 1.  (c) Difference in fiber direction between defined and mapped fiber orientation. The difference is calculated as $err=1-|f_{\text{mapped}}^T f_{\text{defined}}|$.}
\label{fig_reg}
\end{figure}

\begin{figure}[]
\centering
\subfloat{\includegraphics[trim=0 0 0 0,clip,width=0.24\textwidth]{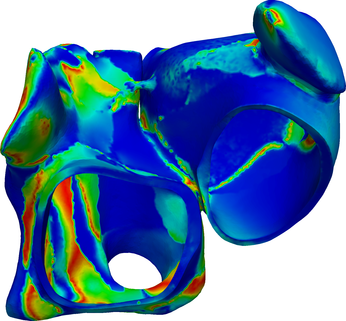}%
\label{fig_atlas_diff_ant_atlas}}
\hfil
\subfloat{\includegraphics[trim=0 0 0 0,clip,width=0.24\textwidth]{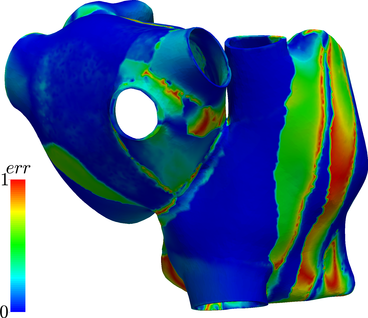}%
\label{fig_atlas_diff_post_atlas}}
\caption{Atlas fibers. Difference in fiber direction between defined and mapped fiber orientation.}
\label{fig_reg_atlas}
\end{figure}

The results of the fiber estimation in Patient 1 and the atlas are shown in Figures \ref{fig_reg} and \ref{fig_reg_atlas}, respectively.
The mapped fibers are overall arranged in quite a similar way as the defined fibers (Figures \ref{fig_fiber_eva_reg} and \ref{fig_fiber_eva_def}). Between the superior vena cava and the inferior vena cava the fibers in the crista terminalis are aligned longitudinally, while the pectinate muscles lay perpendicular to them. Around the vestibulum and the orifices of the veins, the fibers run in circumferential direction. Although the atlas atria has three pulmonary veins compared to four pulmonary veins of Patient 1, the mapped fibers in this area run in circumferential direction around the orifices in both cases (i.e. in case of the atlas registered to Patient 1 as well as in case of the Patient 1 registered to the atlas atria). 

Figure  \ref{fig_eva_diff} shows that the error between mapped and manually defined fibers is small in general (blue color), where larger differences are concentrated at the boundaries between different fiber bundles. For example, the fiber bundle of the crista terminalis is shifted, i.e. it runs a few millimeters further to the left in the mapped case (see Figures \ref{fig_fiber_eva_reg} and \ref{fig_fiber_eva_def}), causing big differences in the error calculation (see Figures \ref{fig_eva_diff}, \ref{fig_reg_atlas}). The error is calculated as $err=1-|f_{\text{mapped}}^T f_{\text{defined}}|$.  Some differences are also visible at the boarders of the circumferential fibers around the veins.

\subsubsection{Electrophysiology}
\begin{figure}[]
\centering
\subfloat[Patient 1 mapped fibers]{\includegraphics[trim=0 0 0 0,clip,width=0.17\textwidth]{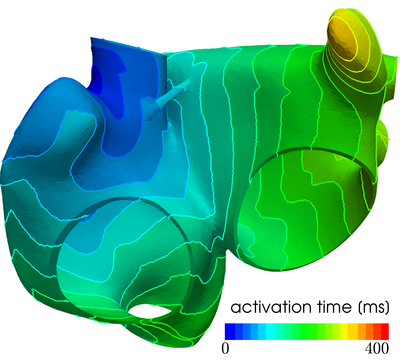}%
\label{activation_time_eva_reg_ant}}
\hfil
\subfloat[Patient 1 defined fibers]{\includegraphics[trim=0 0 0 0,clip,width=0.17\textwidth]{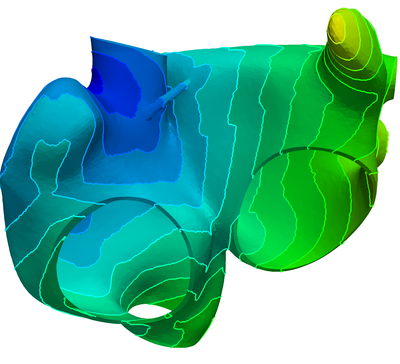}%
\label{activation_time_eva_def_ant}}
\hfil
\subfloat[Patient 1 activation time difference]{
\begin{minipage}{0.5\textwidth}
\centering
                \includegraphics[width=0.34\textwidth]{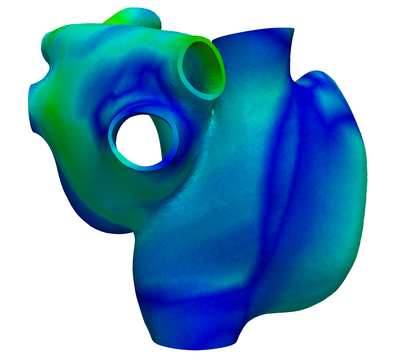}
                \includegraphics[width=0.34\textwidth]{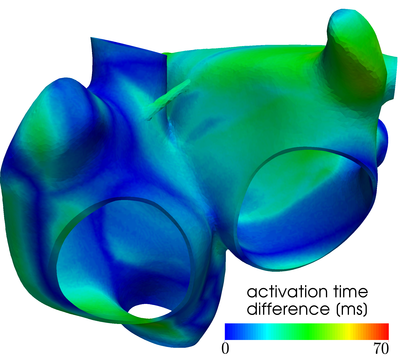}
            \end{minipage}%
\label{actdiff_eva}}
\hfil
\subfloat[Atlas mapped fibers]{\includegraphics[trim=0 0 0 0,clip,width=0.17\textwidth]{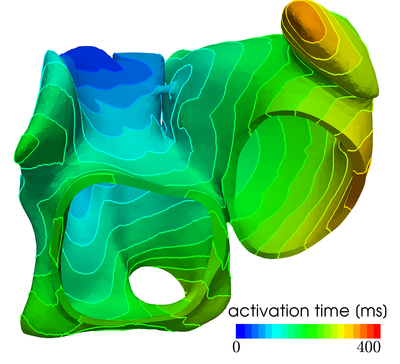}%
\label{activation_time_atlas_reg_ant}}
\hfil
\subfloat[Atlas defined fibers]{\includegraphics[trim=0 0 0 0,clip,width=0.17\textwidth]{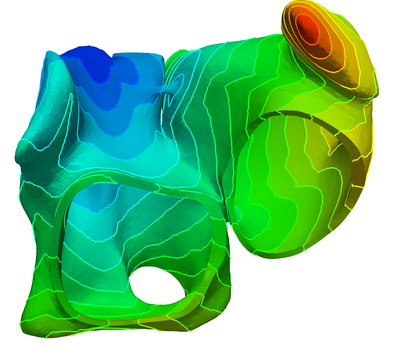}%
\label{activation_time_atlas_def_ant}}
\hfil
\subfloat[Atlas activation time difference]{
\begin{minipage}{0.5\textwidth}
\centering
                \includegraphics[width=0.34\textwidth]{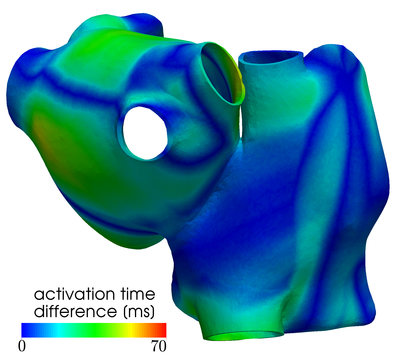}
                \includegraphics[width=0.34\textwidth]{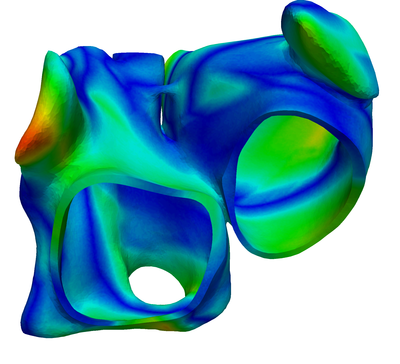}
            \end{minipage}%
\label{actdiff_atlas}}
\hfil
\caption{Results of electrophysiology simulations. Activation times of Patient 1 atria and atlas atria with mapped fibers and defined fibers and activation time differences. The difference is calculated as $\text{diff}=|(\text{activation time})_{\text{mapped}}- (\text{activation time})_{\text{defined}}|$. (a) Activation time with mapped fibers for Patient 1. (b) Activation time with defined fibers for Patient 1. (c) Difference between the activation times with mapped fibers to defined fibers of Patient 1.  (d) Activation time with mapped fibers for the atlas. (e) Activation time with defined fibers for the atlas. (f) Difference between the activation times with mapped fibers to defined fibers of the atlas atria. }
\label{fig_actdiff}
\end{figure}

In Figure \ref{fig_actdiff} the results of the electrophysiological simulations are shown for the atlas atria and Patient 1 atria. Here, the activation times obtained from mapped fibers and manually defined fibers are compared. The overall activation pattern is very similar for both fiber architectures:  the signal travels fast along the crista terminalis, at the same spots the activation travels from the right atrium to the left atrium, and the tip of the left appendage is activated at last. Moreover, in both cases it is clearly visible that the activation of the left atrium occurs through the Bachmann bundle.
Differences in the activation can be seen at the borders of the fiber bundles (compare with Figures \ref{fig_fiber-atlas} and \ref{fig_reg}), as expected due to the aforementioned differences in the fiber orientation.
Characteristics in the activation pattern which appear in the atria with defined fibers also appear in the atria with fibers mapped from it. This can be seen for Patient 1, where the activation sequence is analogous to the atlas atria with defined fibers (compare Figures \ref{activation_time_eva_reg_ant} and \ref{activation_time_atlas_def_ant}).

The transfer of the activation characteristics from the defined to the registered atria are also visible in the activation time. The atlas atria with defined fibers and the Patient 1 atria with mapped fibers take both longer to activate than the Patient 1 atria with defined fibers and the atlas with mapped fibers (see Table \ref{tab_values}).
The maximal difference in activation time is for the atlas atria around \SI{18}{\percent} and for Patient 1 \SI{16}{\percent}. In the atlas atria the region with the biggest difference is the tip of the right appendage, while for Patient 1  it is around the left appendage (see Figure \ \ref{fig_actdiff}). These differences are acceptable due to the fact that they appear in the appendages of the atria, which are less important for the ejection fraction (see next paragraph and Table \ref{tab_values}).

\subsubsection{Mechanical Contraction}
\begin{figure}[htbp!]
\centering
\subfloat[Slice 1]{\includegraphics[width=0.30\textwidth]{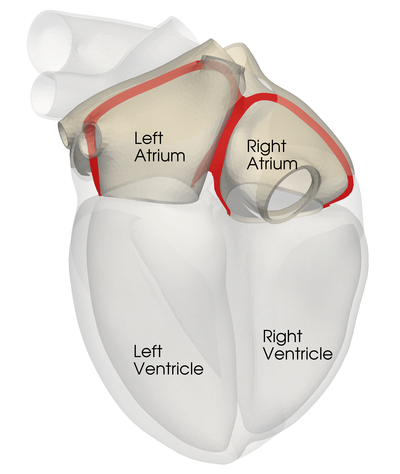}%
\label{fig_eva_slice_4ch}}
\hfill
\subfloat[Slice 2]{
\begin{minipage}[b][][b]{0.4\textwidth}
\centering
\includegraphics[width=0.64\textwidth]{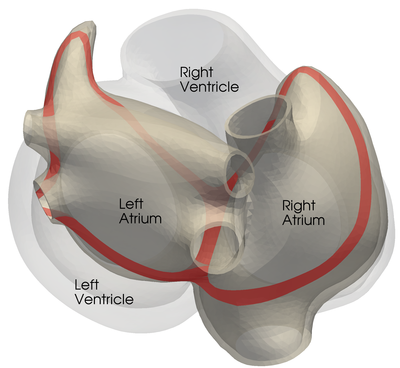}
\includegraphics[width=0.34\textwidth]{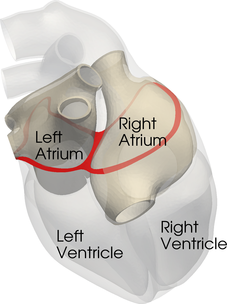}
\label{fig_eva_slice_2ch}
\end{minipage}}%
\caption{Illustrating the slices through the atria which are used to show the contraction. (a) Slice through the atria in the plane of the 4-chamber-view. (b) Slice through the atria in a plane parallel to the heart skeleton.}
\label{fig_slices}
\end{figure}

\begin{figure}[htbp!]
\centering
\subfloat[Mapped fibers]{\includegraphics[width=0.22\textwidth]{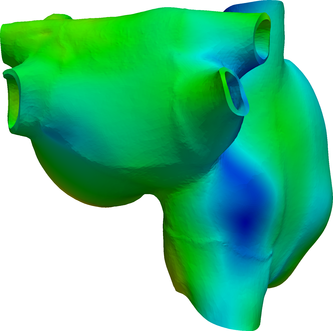}%
\label{fig_disp_eva_reg_left_t320}}
\hfil
\subfloat[Defined fibers]{\includegraphics[width=0.22\textwidth]{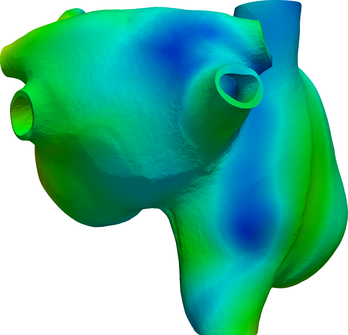}%
\label{fig_disp_eva_def_left_t320}}
\hfil
\subfloat{\includegraphics[width=0.03\textwidth]{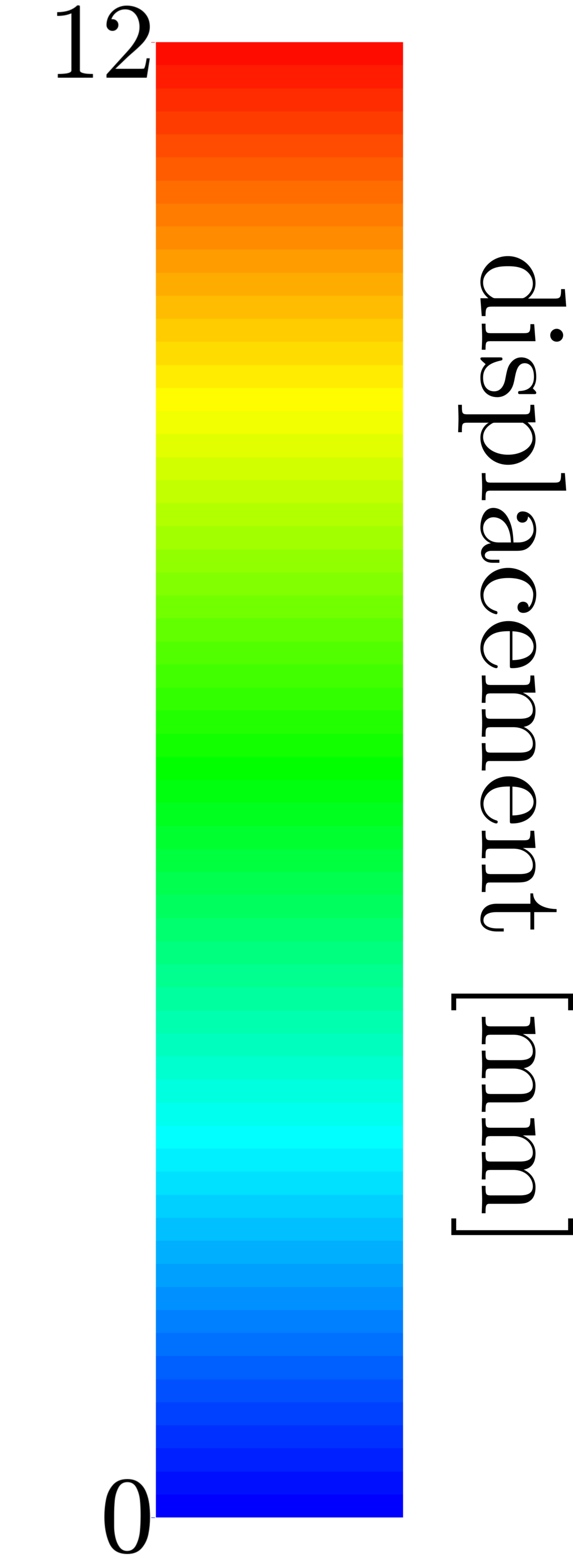}}
\hfil
\addtocounter{subfigure}{-1}
\subfloat[View 1]{\includegraphics[width=0.24\textwidth]{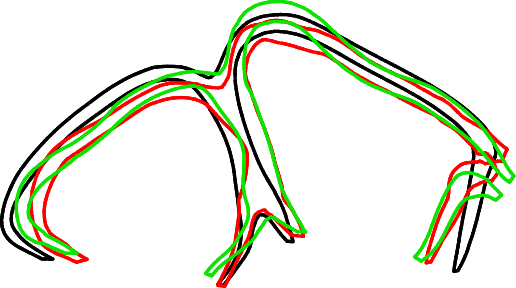}%
\label{fig_disp_eva_slice_4ch_t320}}
\hfil
\subfloat[View 2]{\includegraphics[width=0.24\textwidth]{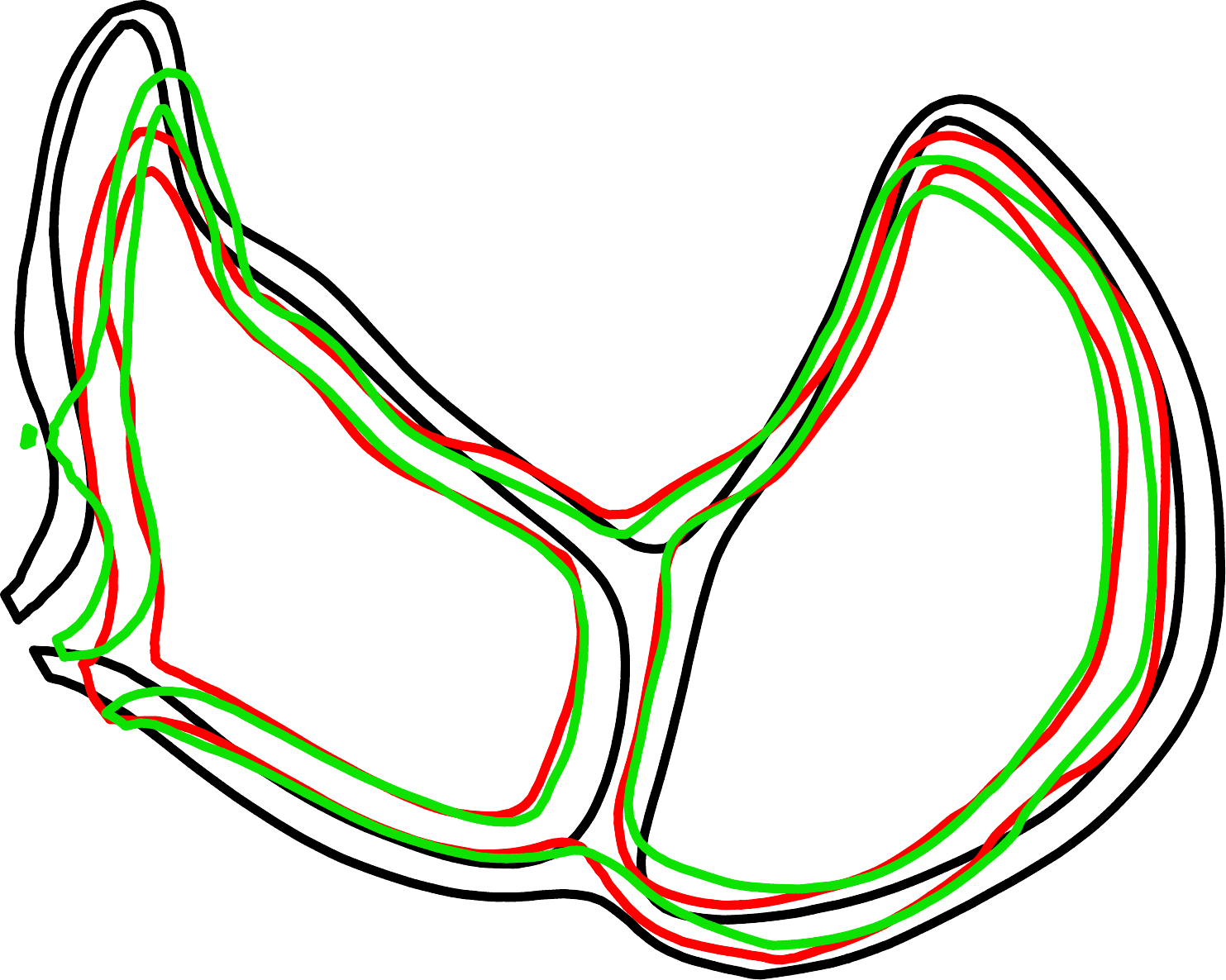}%
\label{fig_disp_eva_slice_2ch_t320}}
\caption{Deformation difference between mapped and defined fibers in the atria for Patient 1 at the time of maximal contraction of the left atrium (time = \SI{0.31}{\second}). (a) and (b) Displacement of the atria for mapped and defined fibers. (c) and (d) Slices through the right and left atrium in the plane shown in Figure \ref{fig_eva_slice_4ch} and Figure \ref{fig_eva_slice_2ch}, respectively. The black contour shows the atrium in the relaxed state, the green contour the contracted atria with defined fibers and the red contour line the contracted atria with mapped fibers.}
\label{fig_eva_disp_left}
\end{figure}

\begin{figure}[htbp!]
\centering
\subfloat[Mapped fibers]{\includegraphics[width=0.22\textwidth]{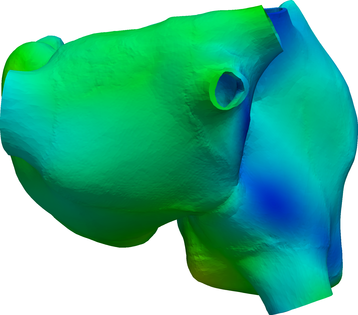}%
\label{fig_disp_atlas_reg_left_t360}}
\hfil
\subfloat[Defined fibers]{\includegraphics[width=0.22\textwidth]{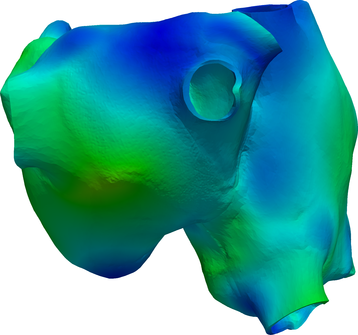}%
\label{fig_disp_atlas_def_left_t360}}
\hfil
\subfloat{\includegraphics[width=0.028\textwidth]{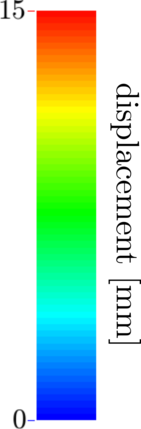}}
\hfil
\addtocounter{subfigure}{-1}
\subfloat[View 1]{\includegraphics[width=0.24\textwidth]{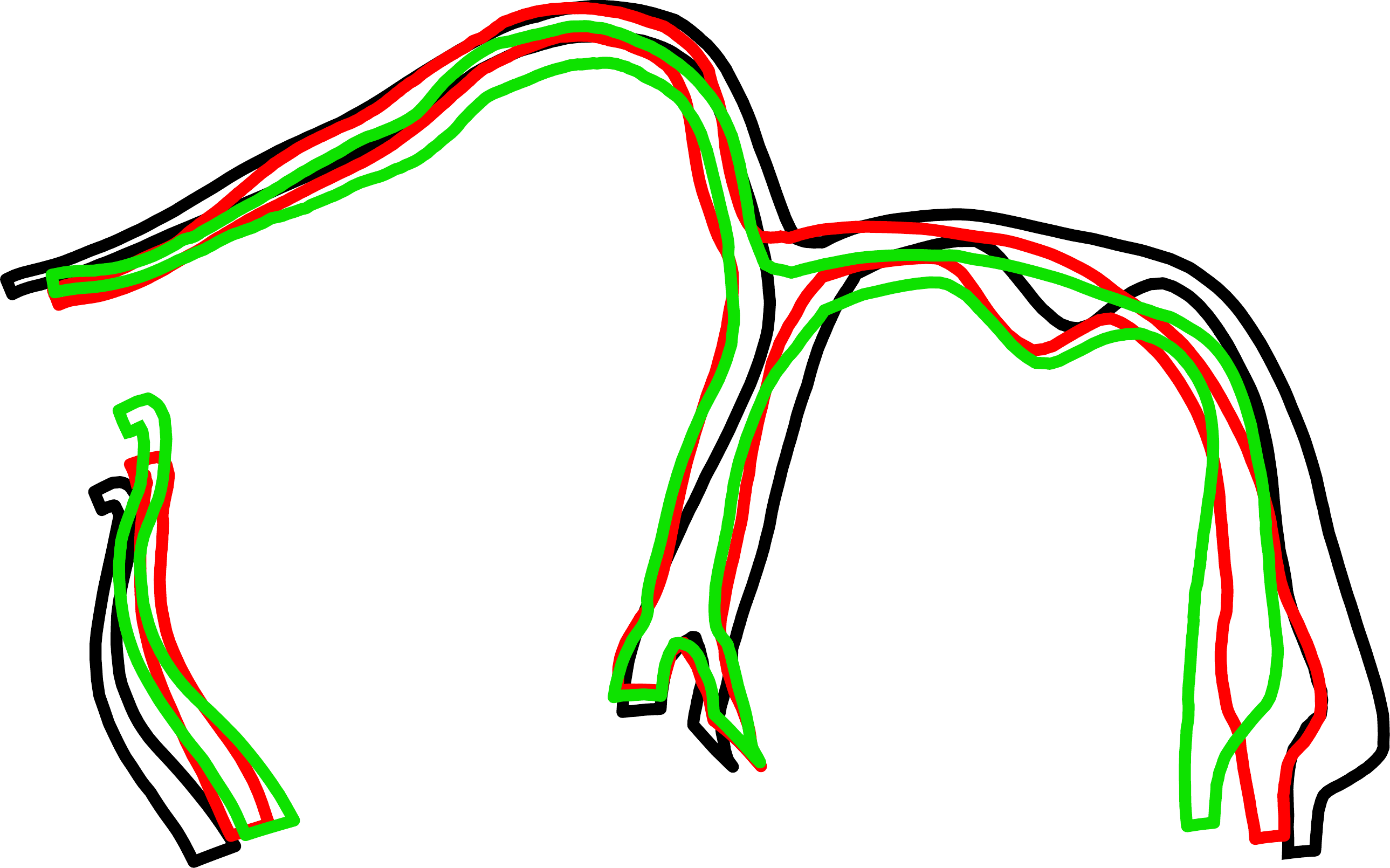}%
\label{fig_disp_atlas_slice_4ch_t360}}
\hfil
\subfloat[View 2]{\includegraphics[width=0.24\textwidth]{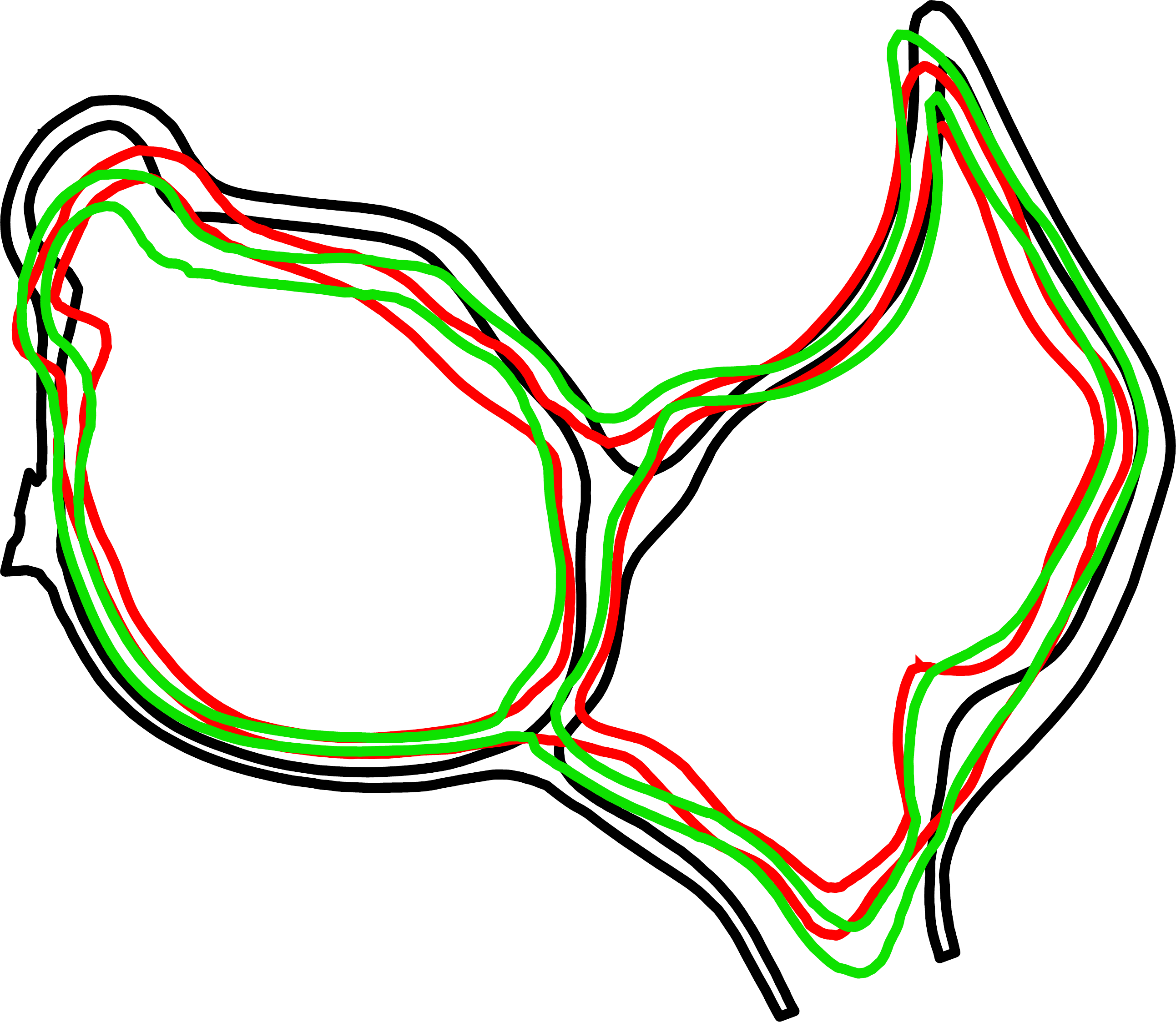}%
\label{fig_disp_atlas_slice_2ch_t360}}
\caption{Deformation difference between mapped and defined fibers in the atria for the atlas atria at the time of maximal contraction of the left atrium (time = \SI{0.35}{\second}). (a) and (b) Displacement of the atria for mapped and defined fibers. (c) and (d) Slices through the right and left atrium in the plane shown in Figure \ref{fig_eva_slice_4ch} and Figure \ref{fig_eva_slice_2ch}, respectively. The black contour shows the atrium in the relaxed state, the green contour the contracted atria with defined fibers and the red contour line the contracted atria with mapped fibers.}
\label{fig_atlas_disp_left}
\end{figure}

\begin{figure}[htbp!]
\centering
\subfloat[Volume Atlas]{\includegraphics[trim=0 0 0 0,clip,width=0.24\textwidth]{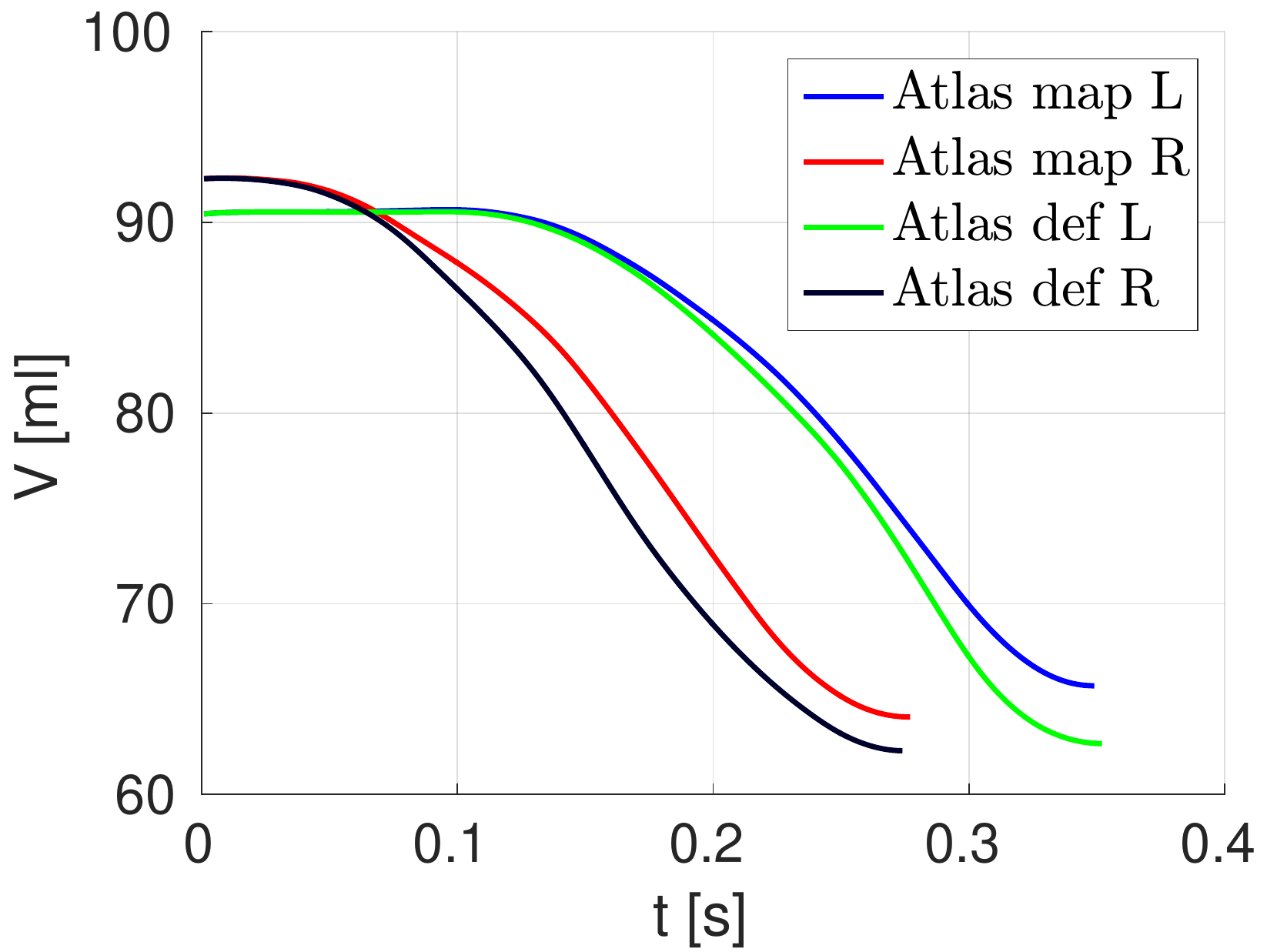}%
\label{fig_atlas_volume}}
\hfil
\subfloat[Pressure Atlas]{\includegraphics[trim=0 0 0 0,clip,width=0.24\textwidth]{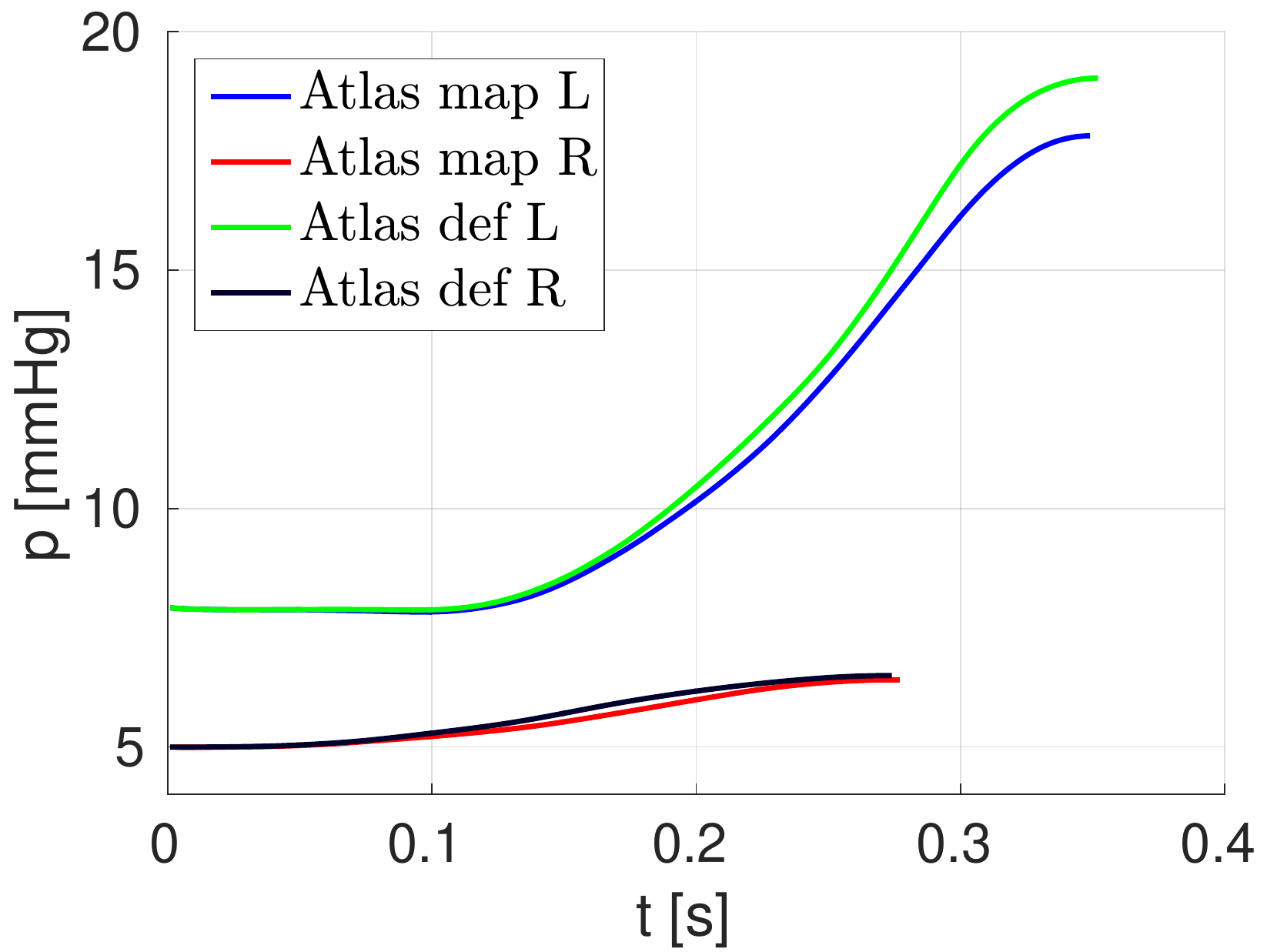}%
\label{fig_atlas_pressure}}
\hfil
\subfloat[Volume Patient 1]{\includegraphics[trim=0 0 0 0,clip,width=0.24\textwidth]{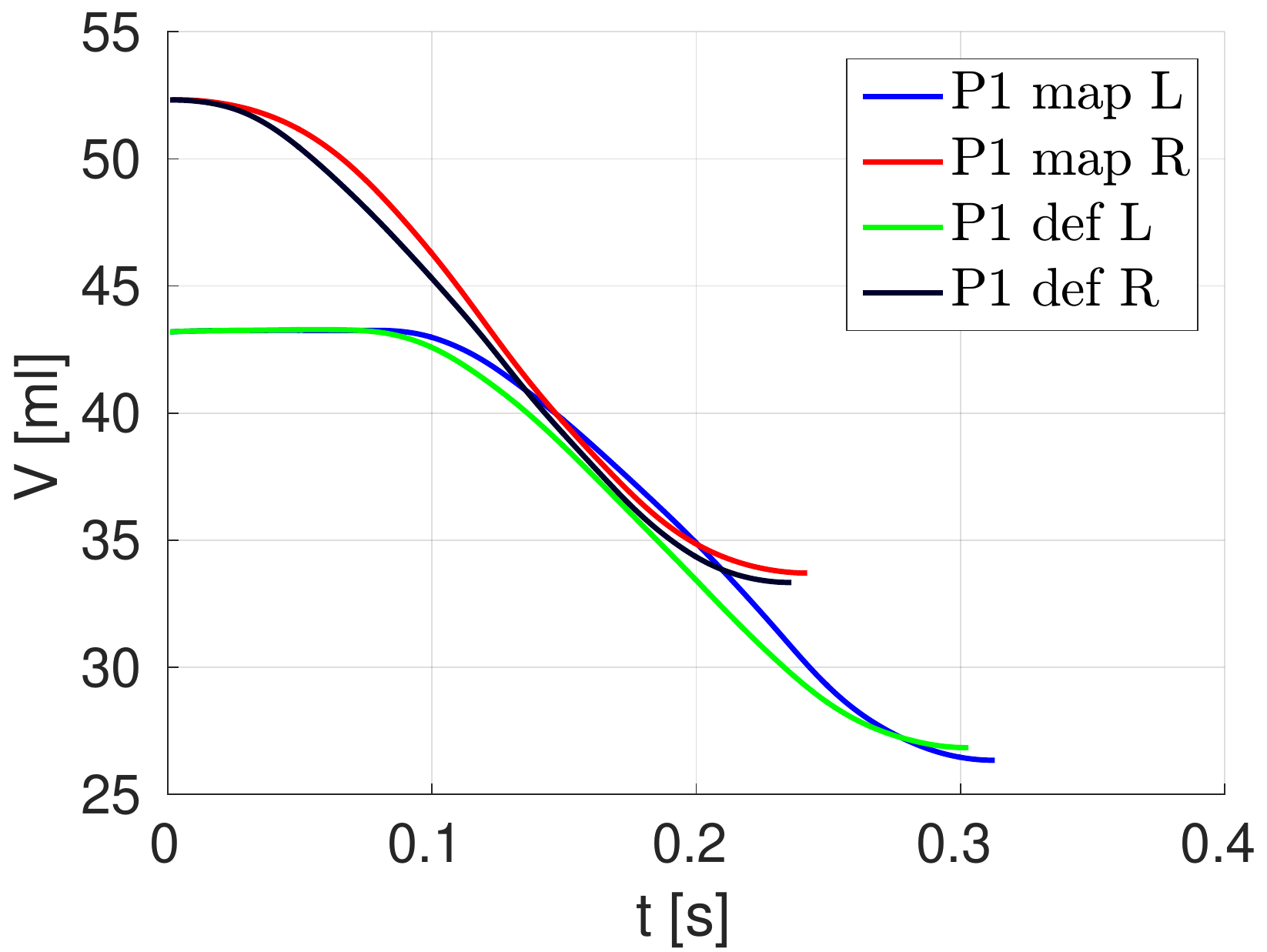}%
\label{fig_eva_volume}}
\hfil
\subfloat[Pressure Patient 1]{\includegraphics[trim=0 0 0 0,clip,width=0.24\textwidth]{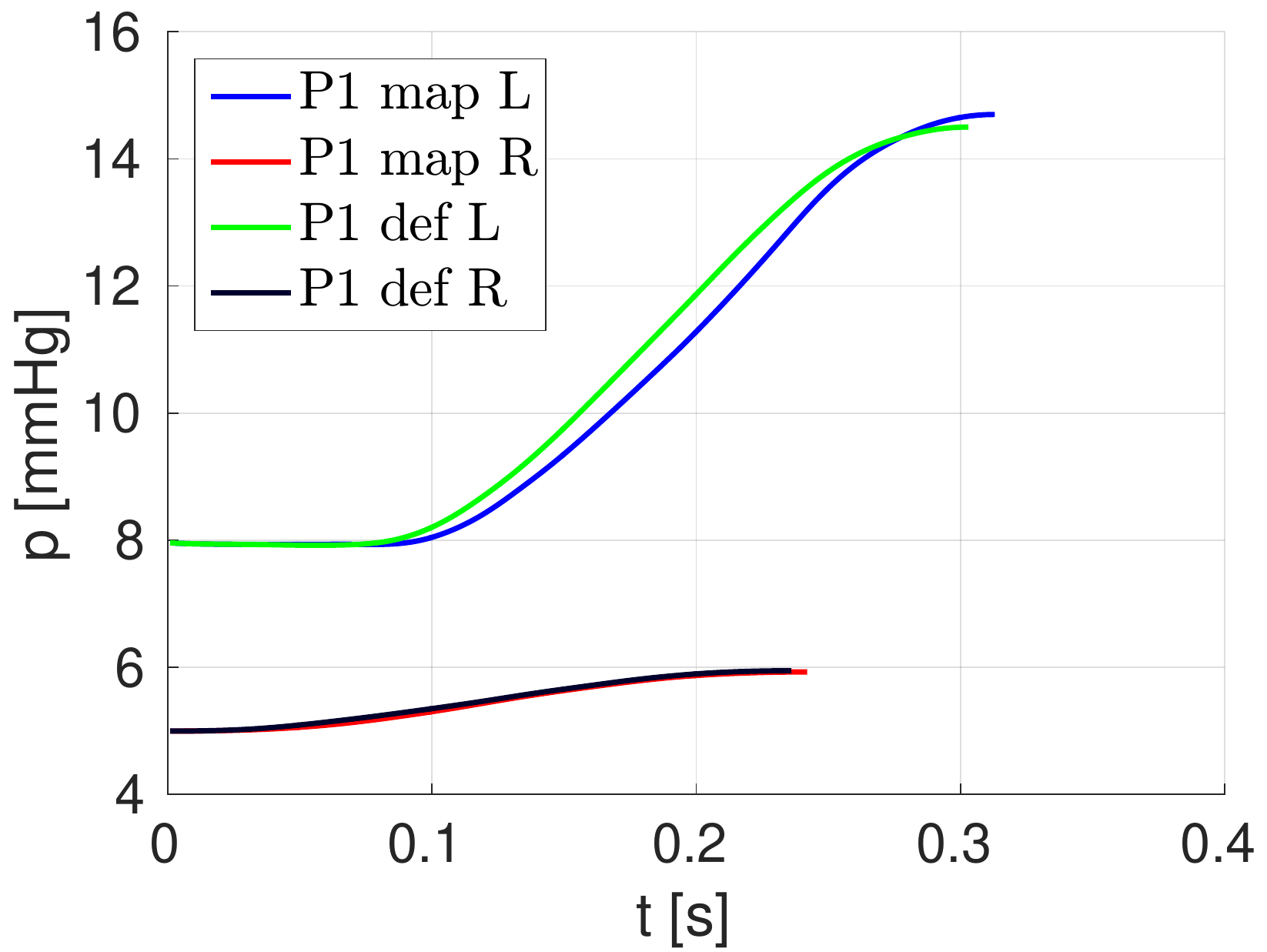}%
\label{fig_eva_pressure}}
\hfil
\caption{Volume (V) and pressure (p) curves plotted over time of the right (R) and left (L) atrium of Patient 1 and the atlas geometry for mapped fibers (map) and defined fibers (def).}
\label{fig_curves}
\end{figure}

The mechanical contraction is coupled with the electrophysiology via the transmembrane potential. The results of the contraction between mapped and defined fibers for the atlas atria and Patient 1 are shown in Figures \ref{fig_eva_disp_left}-\ref{fig_atlas_disp_left}. At the top of each figure the displacements are shown at the time of maximal contraction of the right and left atrium, respectively. The contour plots in the bottom of each figure show the contour of the atria at differently positioned slices through the atria. Figure \ref{fig_slices} shows the position of the slices. Slice 1, visualized in Figure \ref{fig_eva_slice_4ch}, cuts the atria in the plane of the standard 4-chamber-view and Slice 2 (see Figure \ref{fig_eva_slice_2ch}) cuts the atria in a plane parallel to the heart skeleton. We analyzed also the volume curves of the left and right atrium over time until maximal contraction of each chamber (see Figure \ref{fig_curves}). Additionally, in Figure \ref{fig_curves} also the pressures in the 
right and left atrium are plotted. Only small differences in the volume curves for the atlas atria with mapped and defined fibers are visible.
The contraction of the atlas with defined fibers is slightly faster than the contraction of the atlas with 
mapped fibers due to the difference in electrical propagation speed.

\begin{figure}[htbp!]
\centering
\subfloat[Right Atrium]{\includegraphics[trim=0 0 0 0,clip,width=0.24\textwidth]{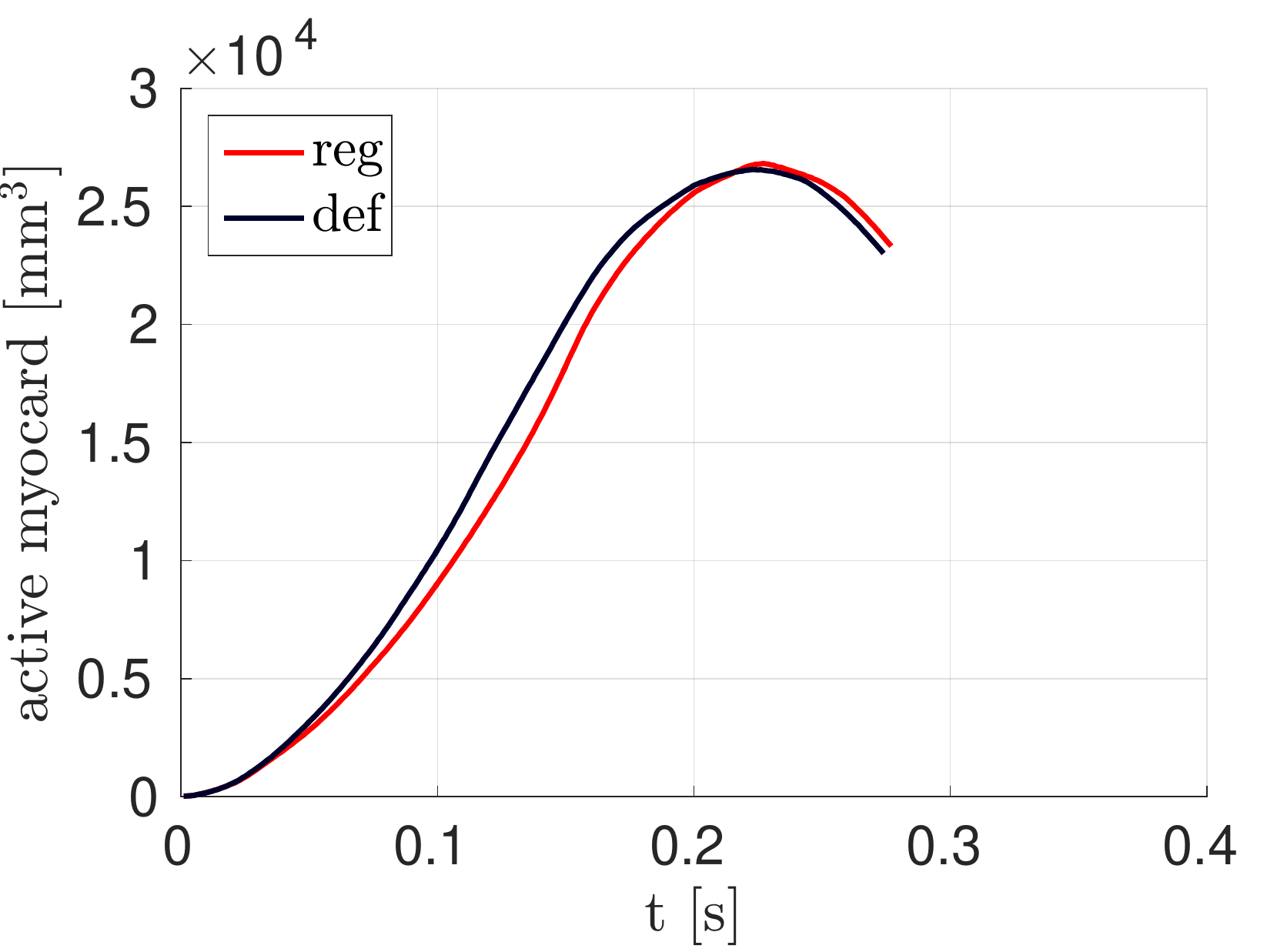}%
\label{fig_atlas_acttissue_right}}
\hfil
\subfloat[Left Atrium]{\includegraphics[trim=0 0 0 0,clip,width=0.24\textwidth]{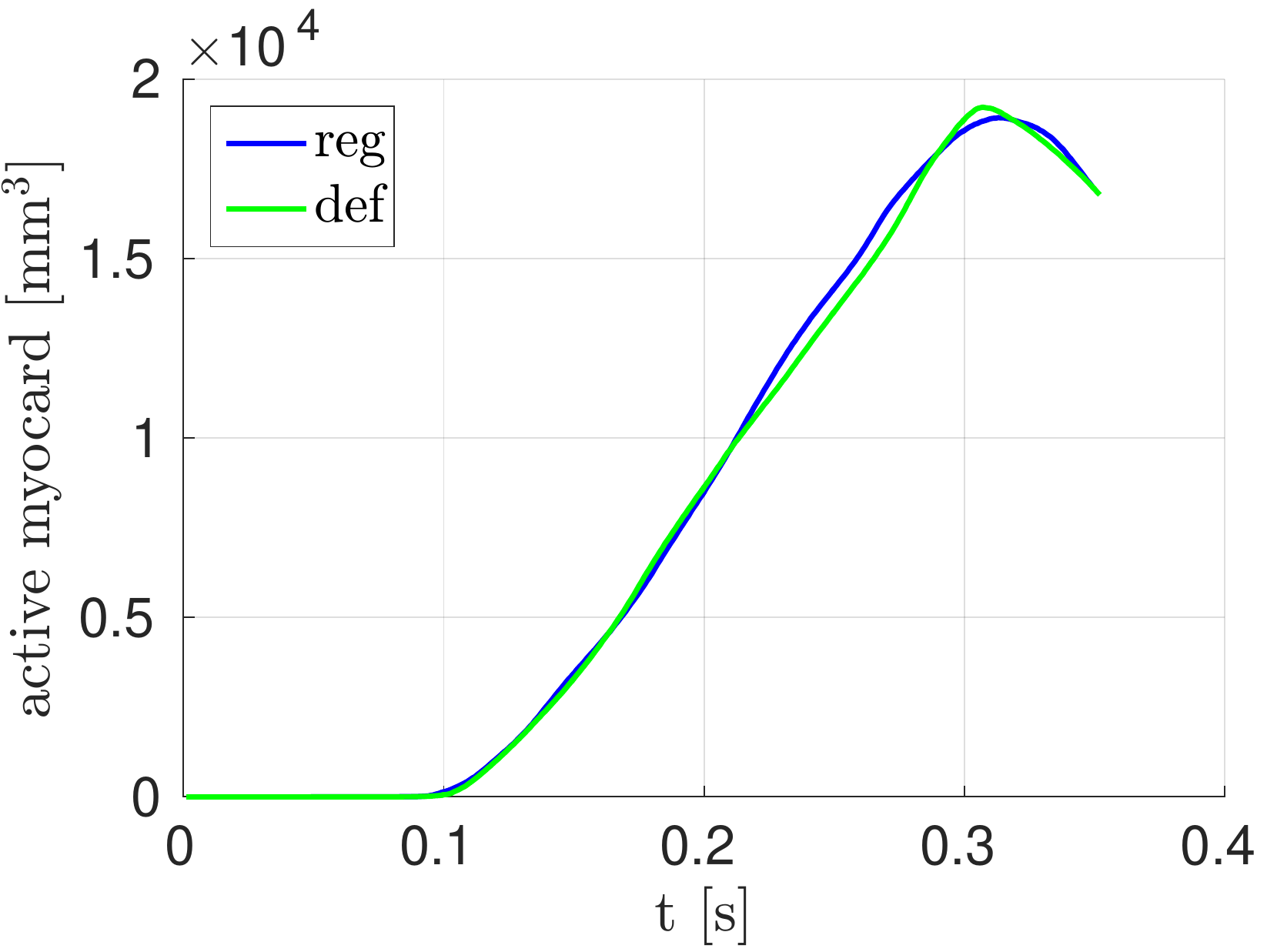}%
\label{fig_atlas_acttissue_left}}
\hfil
\caption{Volume of active contracting tissue over time for the atlas atria with defined and mapped fibers}
\label{fig_acttissue}
\end{figure}

To see the influence of the different activation we plotted the volume of actively contracting tissue, i.e. the tissue where the action potential is above a threshold, over time. In the plot the active tissue is compared for the atlas atria with defined fibers to the atlas atria with mapped fibers. Figure \ref{fig_atlas_acttissue_right} shows the right atrium and Figure \ref{fig_atlas_acttissue_left} left atrium. It is visible in the plot that for the right atrium slightly more tissue is active in the atrium with defined fibers than with mapped fibers at the same time. This is consistent with the faster decrease in the volume curve (see Figure \ref{fig_atlas_volume}). 
The slight shift of the fibers of the crista terminalis on the posterior side of the right atrium (see Paragraph \ref{sec_fibers}) in the mapped case, leads to fibers in the thickened wall of the crest not running along the crest. This is the reason for the folding near the terminal crest during contraction of the right atrium (see Figure \ref{fig_atlas_disp_left}). Especially in Figure \ref{fig_disp_atlas_slice_2ch_t360} the folding  of the terminal crest is visible. However, both mapped and defined fibers fold in the regions of fiber orientation change. Differences in the displacement of the left atrial wall of the atlas atria during contraction exist near characteristic shapes, which only exist in individual atria, for example, the pronounced buckle in the inferior wall of the right appendage and the distinctive shape of the tip of the right appendage. Since these characteristics are different and only present in individual atria, the registration process, especially the fiber interpolation, cannot map 
the correct fiber directions in this regions. However, they are also not known from anatomical studies. In Patient 1 we have a very smooth shaped atria. The difference in displacements during 
contraction is smaller between mapped and defined fibers (see Figures \ref{fig_eva_disp_left} and \ref{fig_atlas_disp_left}). Also the volume and pressure curves, the ejection fraction and the time of maximal contraction are similar. Only the displacement of the pulmonary veins, the caval veins and the left appendage, differ slightly. For the atlas atria and the 
atria of Patient 1 characteristic values of the activation and contraction are listed in Table \ref{tab_values}.

\begin{figure}[htbp!]
\centering
\subfloat[Volume Left Atrium]{\includegraphics[trim=0 0 0 0,clip,width=0.24\textwidth]{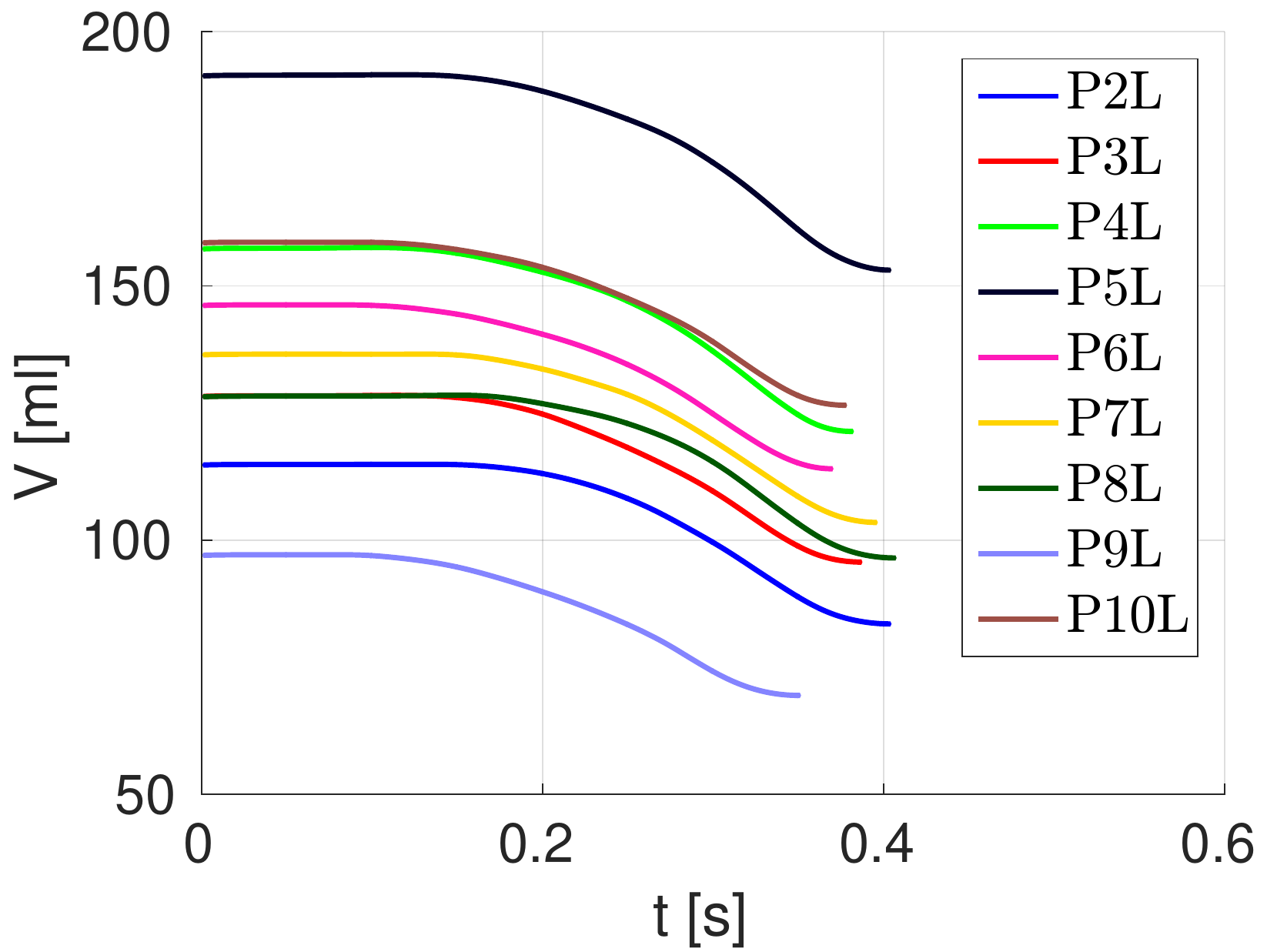}%
\label{fig_patient_left_volume}}
\hfil
\subfloat[Pressure Left Atrium]{\includegraphics[trim=0 0 0 0,clip,width=0.24\textwidth]{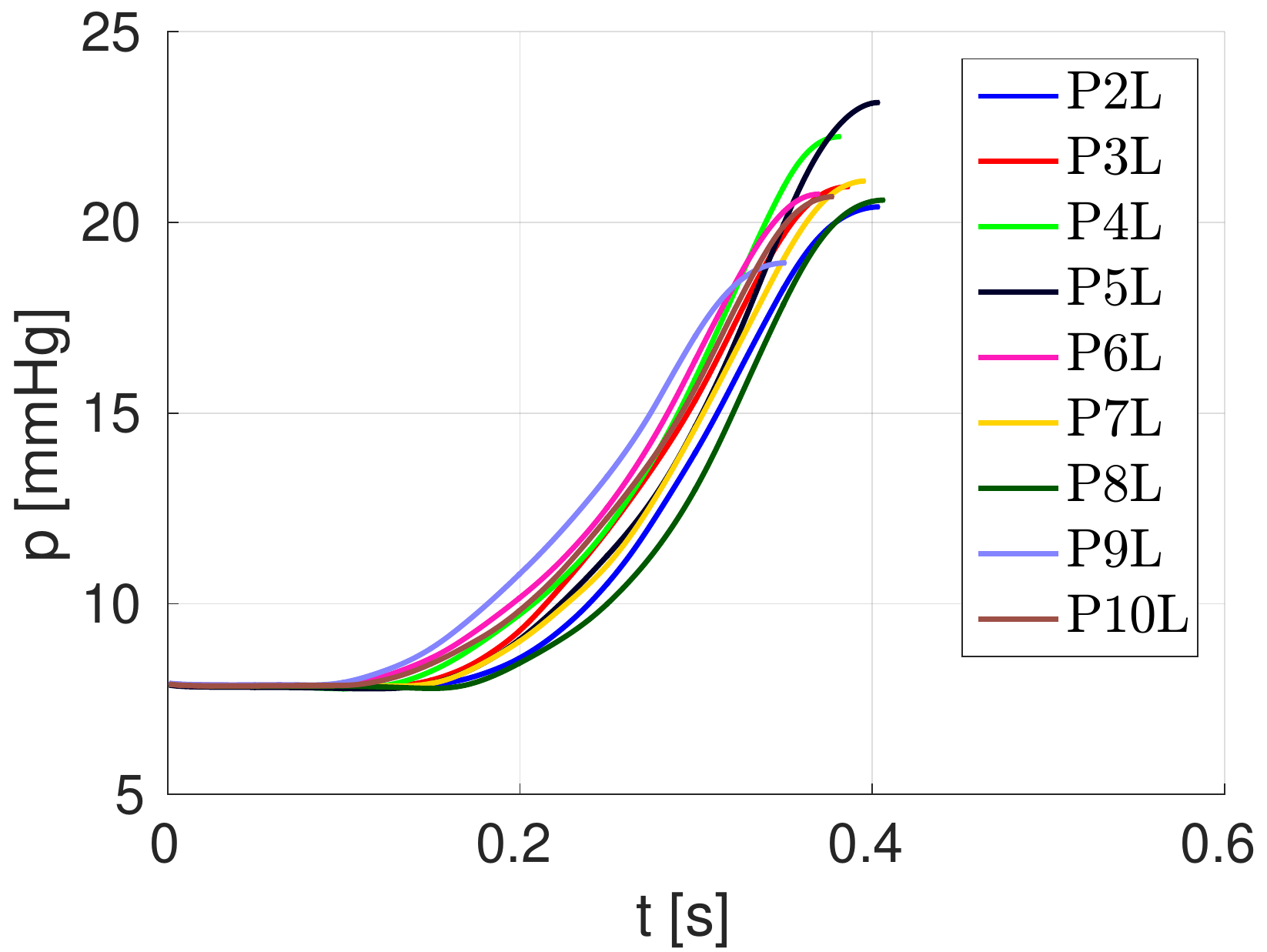}%
\label{fig_patient_left_pressure}}
\hfil
\subfloat[Volume Right Atrium]{\includegraphics[trim=0 0 0 0,clip,width=0.24\textwidth]{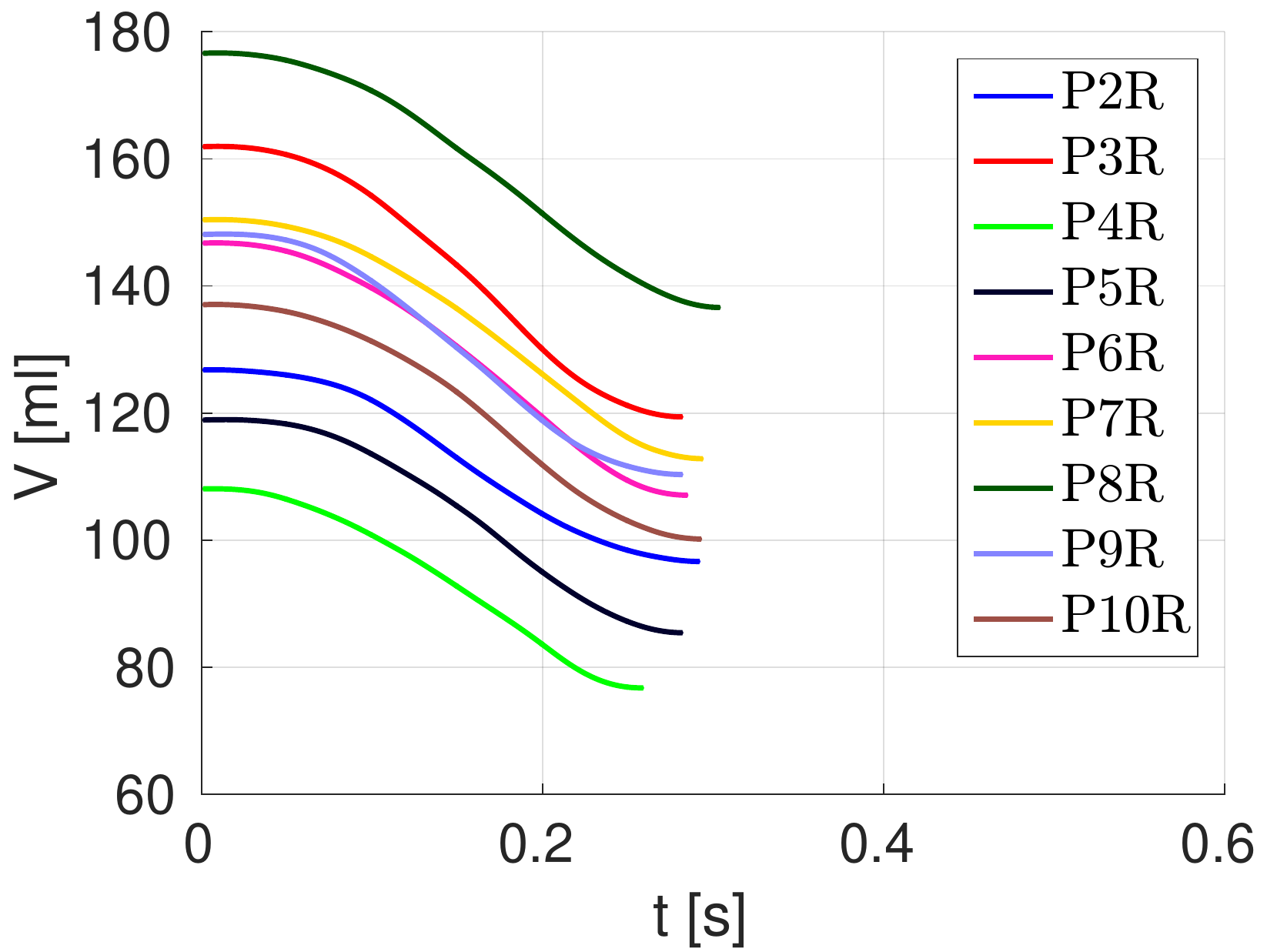}%
\label{fig_patient_right_volume}}
\hfil
\subfloat[Pressure Right Atrium]{\includegraphics[trim=0 0 0 0,clip,width=0.24\textwidth]{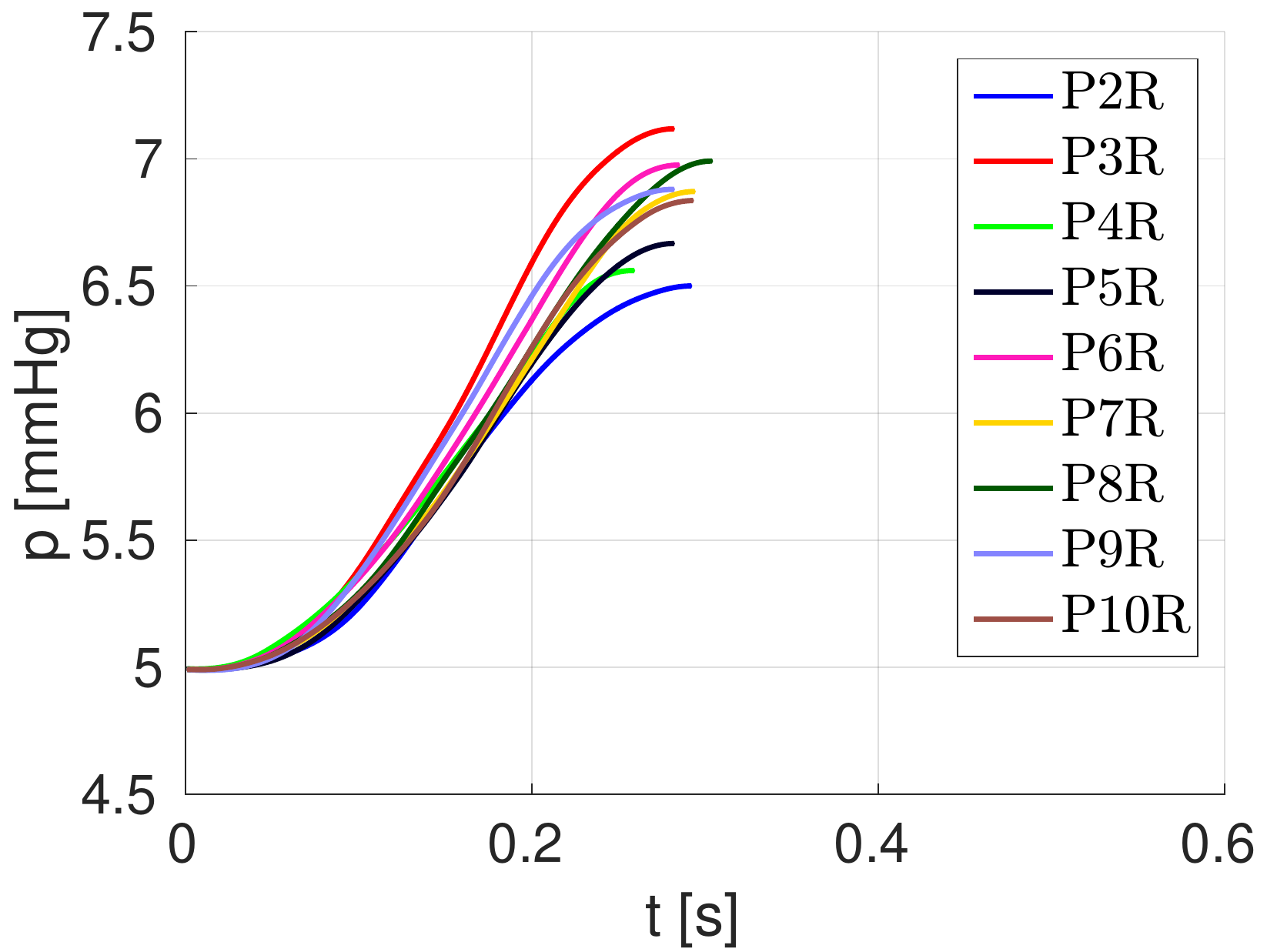}%
\label{fig_patient_right_pressure}}
\caption{Volume (V) and pressure (p) curves plotted over time of the right (R) and left (L) atrium for Patient 2-10.}
\label{fig_patient_curves}
\end{figure}

\begin{table*}[htbp!]
\renewcommand{\arraystretch}{1.3}
%if using array.sty, it might be a good idea to tweak the value of
%\extrarowheight as needed to properly center the text within the cells
\caption{Summary of electrophysiology and contraction outputs for defined and mapped fiber orientations in atlas  and Patient 1 geometries.}
\label{tab_values}
\centering
\sisetup{round-mode= places,round-precision = 1}
% Some packages, such as MDW tools, offer better commands for making tables
% than the plain LaTeX2e tabular which is used here.
\begin{tabular}{|cl||cccccc|}
\hline
 & & Activation Time [s] & Start Activation [s] & Max/Min Volume [ml] & EF [\%] & Max Pressure [mmHg] & Time Min. Volume [s]\\
 \hline
 \hline
 Left &&&&&&&\\
 & Atlas map. & \num[round-precision = 3]{0.338} & \num[round-precision = 3]{0.098}&  \num{90.237613} / \num{65.7015717816899} & \num{27.5732948824884}  & \num{17.8143418467805}  & \num[round-precision = 3]{0.349}  \\
 &Atlas def. & \num[round-precision = 3]{0.376} & \num[round-precision = 3]{0.09}& \num{90.237613} / \num{62.6711304704444} & \num{30.8908730610759} & \num{19.026517701898}  & \num[round-precision = 3]{0.352} \\
 &Patient 1 map. & \num[round-precision = 3]{0.317} & \num[round-precision = 3]{0.076}& \num{43.08893} / \num{26.3452440603676} & \num{39.5473506623582} & \num{14.6974119280259}  & \num[round-precision = 3]{0.313}\\
 &Patient 1 def. & \num[round-precision = 3]{0.288} & \num[round-precision = 3]{0.062}& \num{43.08893} / \num{26.8412097214347} & \num{38.2789084919675} & \num{14.4990183282687}  & \num[round-precision = 3]{0.303} \\ 
 \hline
 Right &&&&&&&\\
 &Atlas map.  &\num[round-precision = 3]{0.262} & \num{0} &\num{92.1740} / \num{64.0666398285335}  & \num{31.0299927373123} & \num{6.4061280870252}  & \num[round-precision = 3]{0.277} \\
 &Atlas def. & \num[round-precision = 3]{0.277}&\num{0} & \num{92.1740} / \num{62.2938025487618} & \num{33.1849331976425}  & \num{6.49477659640426} & \num[round-precision = 3]{0.274}\\
 &Patient 1 map. &\num[round-precision = 3]{0.212} &\num{0} & \num{52.2500} / \num{33.708274894174} &  \num{36.0112309123391} & \num{5.92783762174759}  & \num[round-precision = 3]{0.242}\\
 &Patient 1 def. & \num[round-precision = 3]{0.212} &\num{0} & \num{52.2500} / \num{33.3374851816822} &  \num{36.7043375625503} & \num{5.94637927986155}   & \num[round-precision = 3]{0.236} \\
\hline
\end{tabular}
\end{table*}

\subsection{Performance Study}

\begin{figure*}[htbp!]
\centering
%\subfloat[Atlas]{\includegraphics[trim=0 0 0 0,clip,width=0.2\textwidth]{fiber_atlas_def_post}}
%\hfil
\subfloat[Patient 2]{\includegraphics[trim=0 0 0 0,clip,width=0.2\textwidth]{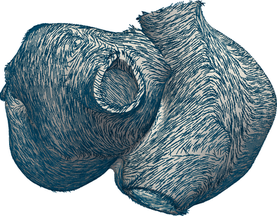}}
\hfil
\subfloat[Patient 3]{\includegraphics[trim=0 0 0 0,clip,width=0.2\textwidth]{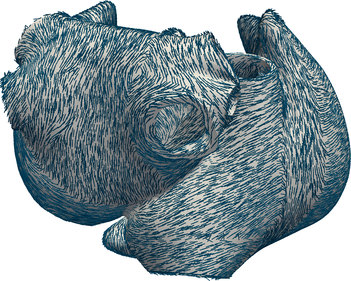}}
\hfil
\subfloat[Patient 4]{\includegraphics[trim=0 0 0 0,clip,width=0.2\textwidth]{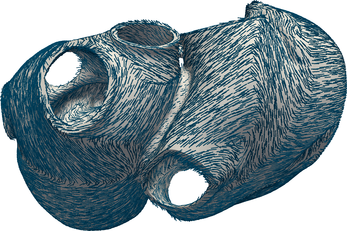}}
\hfil
\subfloat[Patient 5]{\includegraphics[trim=0 0 0 0,clip,width=0.2\textwidth]{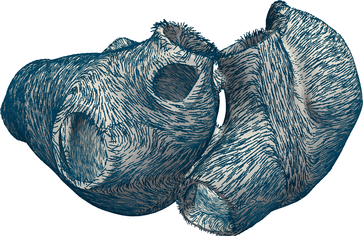}\label{fig_fiber_heart4}}
\hfil
\subfloat[Patient 6]{\includegraphics[trim=0 0 0 0,clip,width=0.2\textwidth]{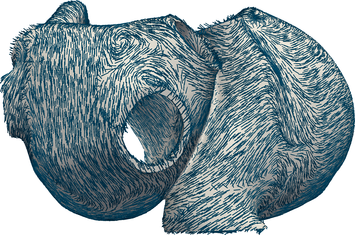}}
\hfil
\subfloat[Patient 7]{\includegraphics[trim=0 0 0 0,clip,width=0.2\textwidth]{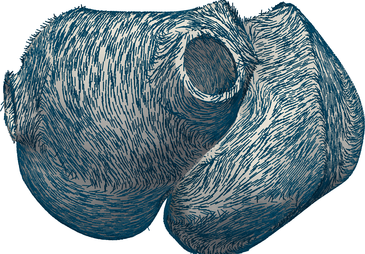}}
\hfil
\subfloat[Patient 8]{\includegraphics[trim=0 0 0 0,clip,width=0.2\textwidth]{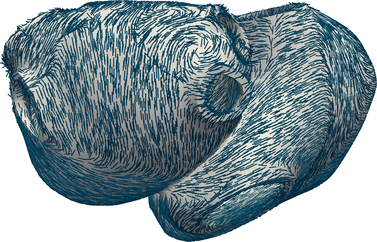}}
\hfil
\subfloat[Patient 9]{\includegraphics[trim=0 0 0 0,clip,width=0.2\textwidth]{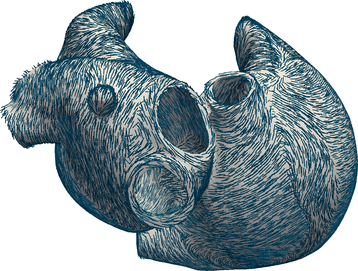}\label{fig_fiber_heart8}}
\hfil
\subfloat[Patient 10]{\includegraphics[trim=0 0 0 0,clip,width=0.2\textwidth]{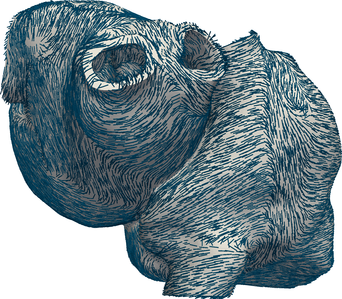}}
\hfil
\caption{Mapped fiber orientation for Patient 2-10.}
\label{fig_estimated_fibers}
\end{figure*}

In Figure \ref{fig_estimated_fibers} all patient's atria with mapped fibers are shown. Although, the geometry of the atria has a different shape in every patient, the registration is able to deform the atlas shape to the correct patient's shape. Additionally, the fiber architecture of the atria is well defined, i.e. every atria has the same fibers bundles which run in the same directions, for example the crista terminalis, the pectinate muscles, the circumferential vestibule fibers etc. Unusual shapes as in Patient 9 are handled well (see Figure \ref{fig_fiber_heart8}). Although the geometry has a bump on the left atrium the fiber direction around it is smooth and also the pump is equipped with fibers. Patient 5, which has an additional right pulmonary vein orifice has physiological fiber orientation in the region of the pulmonary orifices (see Figure \ref{fig_fiber_heart4}). It is important to remark that no unphysiological fiber orientations were detected.

\begin{figure*}[htbp!]
\centering
%\subfloat[Atlas]{\includegraphics[trim=0 0 0 0,clip,width=0.25\textwidth]{act_atlas_ant}}
%\hfil
%\subfloat[Patient 1]{\includegraphics[trim=0 0 0 0,clip,width=0.25\textwidth]{act_eva_ant}}
%\hfil
\subfloat[Patient 2]{\includegraphics[trim=0 0 0 0,clip,width=0.2\textwidth]{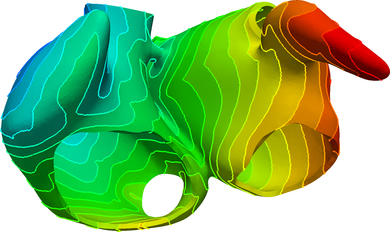}\label{fig_act_heart1}}
\hfil
\subfloat[Patient 3]{\includegraphics[trim=0 0 0 0,clip,width=0.2\textwidth]{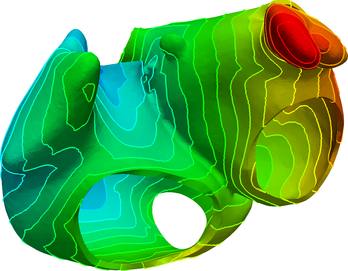}}
\hfil
\subfloat[Patient 4]{\includegraphics[trim=0 0 0 0,clip,width=0.2\textwidth]{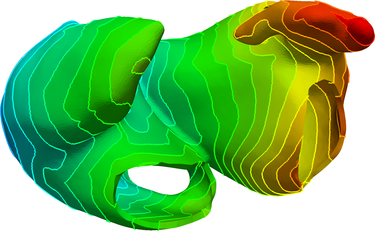}}
\hfil
\subfloat[Patient 5]{\includegraphics[trim=0 0 0 0,clip,width=0.2\textwidth]{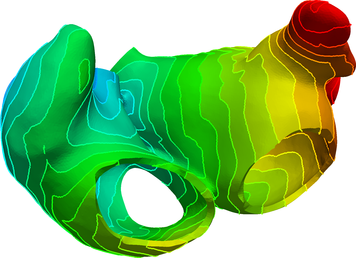}\label{fig_act_heart4}}
\hfil
\subfloat[Patient 6]{\includegraphics[trim=0 0 0 0,clip,width=0.2\textwidth]{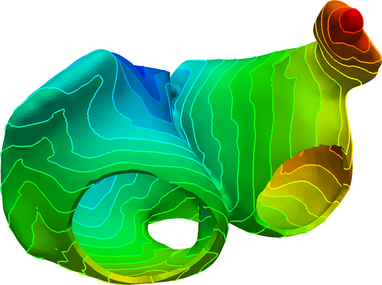}\label{fig_act_heart5}}\\
\hfil
\subfloat[Patient 7]{\includegraphics[trim=0 0 0 0,clip,width=0.2\textwidth]{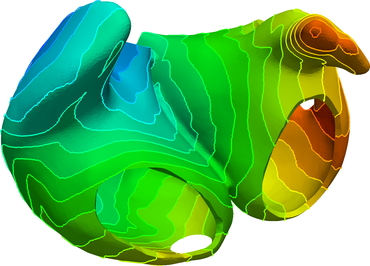}}
\hfil
\subfloat[Patient 8]{\includegraphics[trim=0 0 0 0,clip,width=0.2\textwidth]{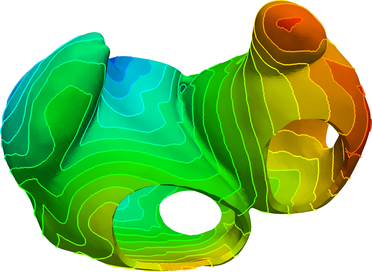}\label{fig_act_heart7}}
\hfil
\subfloat[Patient 9]{\includegraphics[trim=0 0 0 0,clip,width=0.2\textwidth]{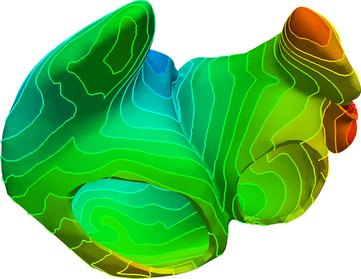}\label{fig_act_heart8}}
\hfil
\subfloat[Patient 10]{\includegraphics[trim=0 0 0 0,clip,width=0.2\textwidth]{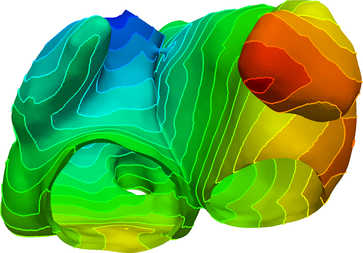}\label{fig_act_heart9}}
\hfil
\subfloat{\includegraphics[trim=0 0 0 0,clip,width=0.100\textwidth]{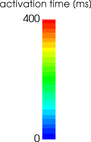}}
\caption{Activation times for the patients' atria Patient 2 to Patient 10 with mapped fiber orientation.}
\label{fig_acttime_all}
\end{figure*}

The results of the electrophysiological simulations are shown in Figure \ref{fig_acttime_all}, where the activation patterns of all patients' atria are shown. It is visible, that the characteristics of the electrophysiological activation pattern which appear in the atlas atria also appear in the patients' atria. One example is the increased propagation speed due to circumferential vestibule fibers in the posterior wall of the left atrium. When the signal reaches the fibers around the vestibule on the posterior left atrial wall, the activation in circumferential direction around the vestibule increases, which is recognizable due to a more or less distinct triangular shape of the activation pattern in this region in each patient's atria (see Figure \ref{fig_acttime_all}). The signal travels from the right atrium to the left atrium through the Bachmann bundle if the septum is small and the distance between Bachmann bundle and 
other interatrial 
connections is big (see Figures \ref{fig_act_heart1}, \ref{fig_act_heart4}, and \ref{fig_act_heart9}). In other atria, the activation through the Bachmann bundle is less important, since other interatrial connections are nearby (see e.g. Figure \ref{fig_act_heart5}, \ref{fig_act_heart7}, and \ref{fig_act_heart8}). Both scenarios are physiological \cite{markides_characterization_2003}. The region which is activated last is the left atrial appendage. However, the overall activation time depends of course on the size and shape of the atria.

\begin{figure}[htbp!]
\centering
\subfloat[Patient 2]{\includegraphics[width=0.15\textwidth]{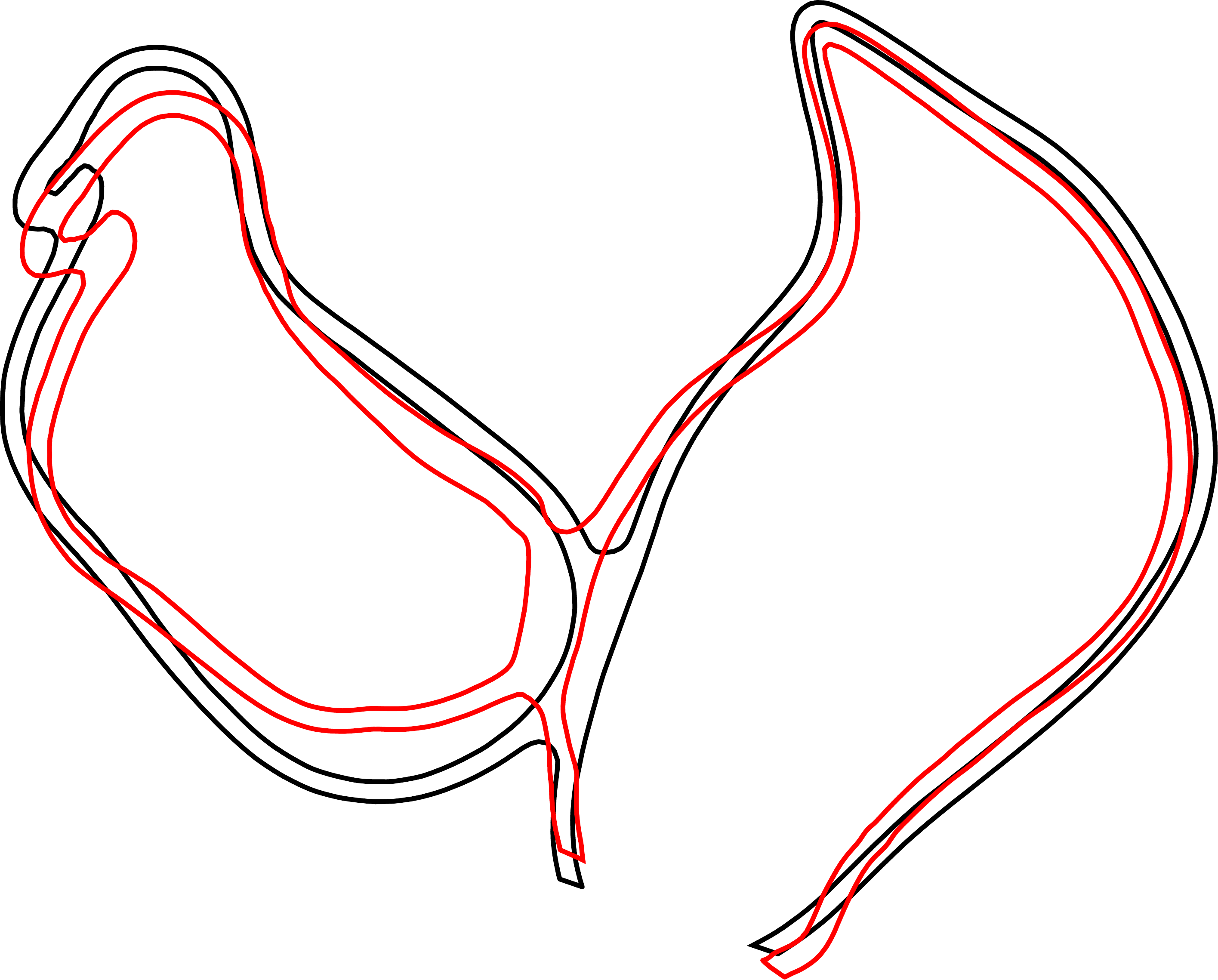}}%
\hfil
\subfloat[Patient 3]{\includegraphics[width=0.15\textwidth]{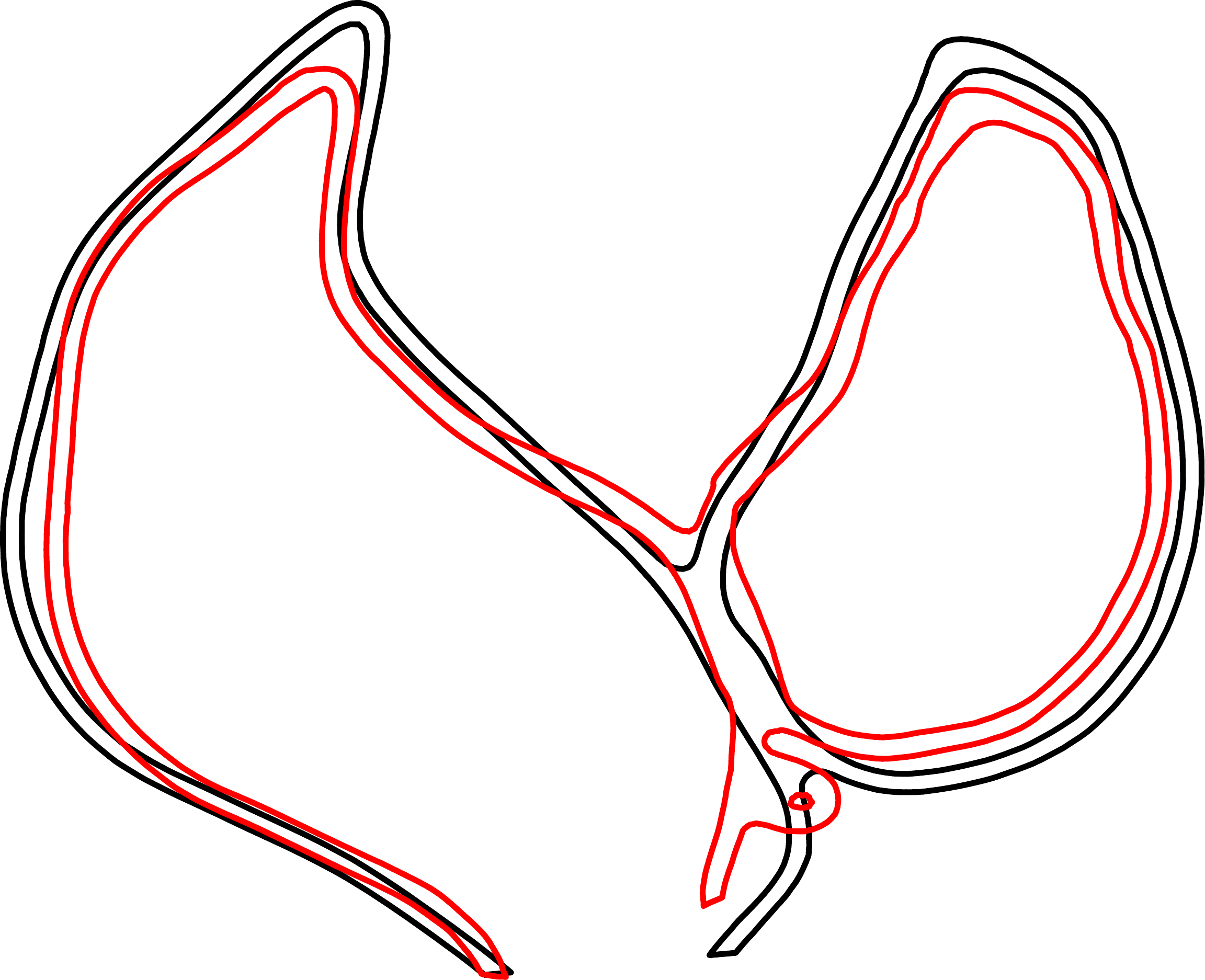}}%
\hfil
\subfloat[Patient 4]{\includegraphics[width=0.15\textwidth]{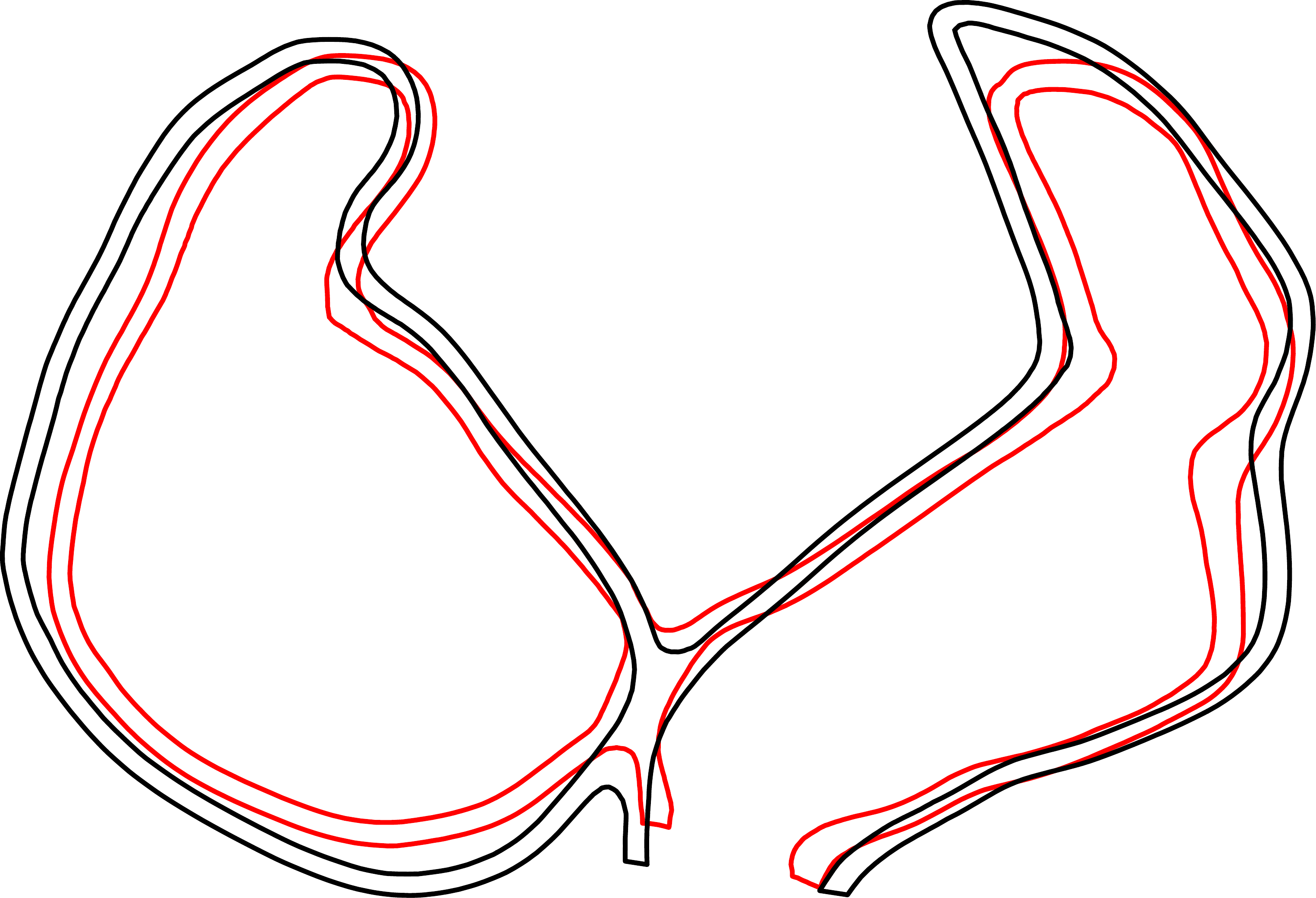}}%
\hfil
\subfloat[Patient 5]{\includegraphics[width=0.15\textwidth]{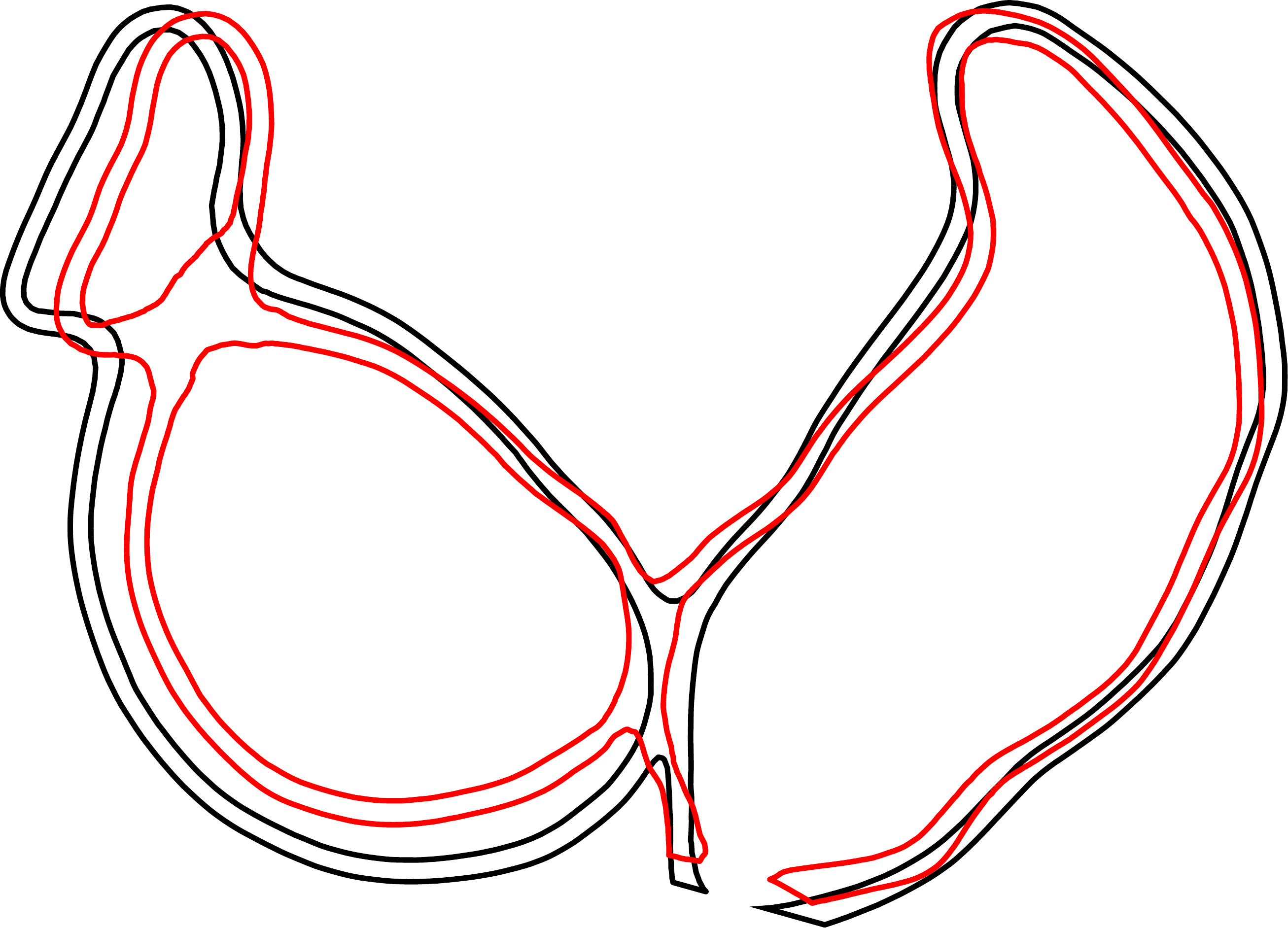}}%
\hfil
\subfloat[Patient 6]{\includegraphics[width=0.15\textwidth]{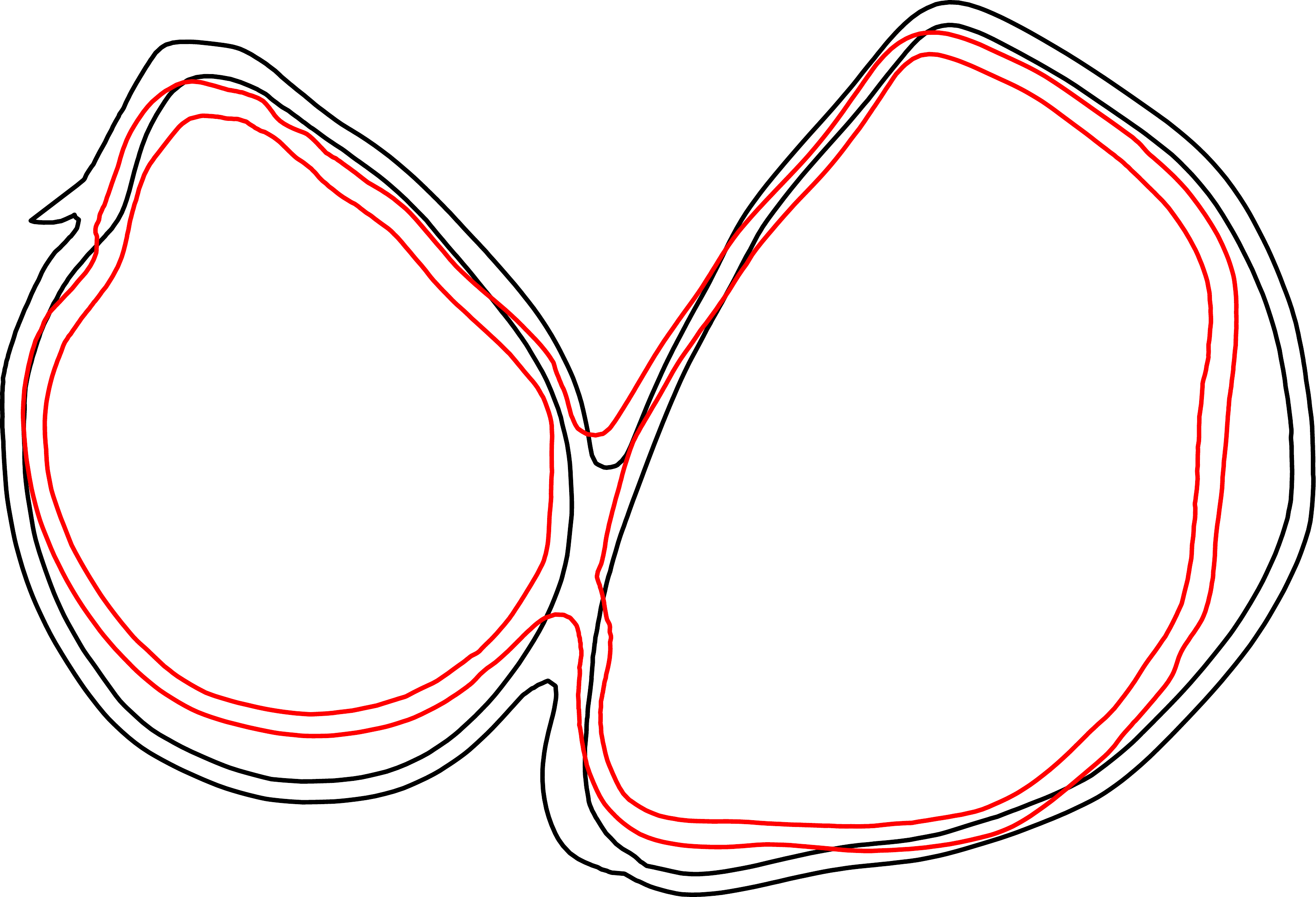}}%
\hfil
\subfloat[Patient 7]{\includegraphics[width=0.15\textwidth]{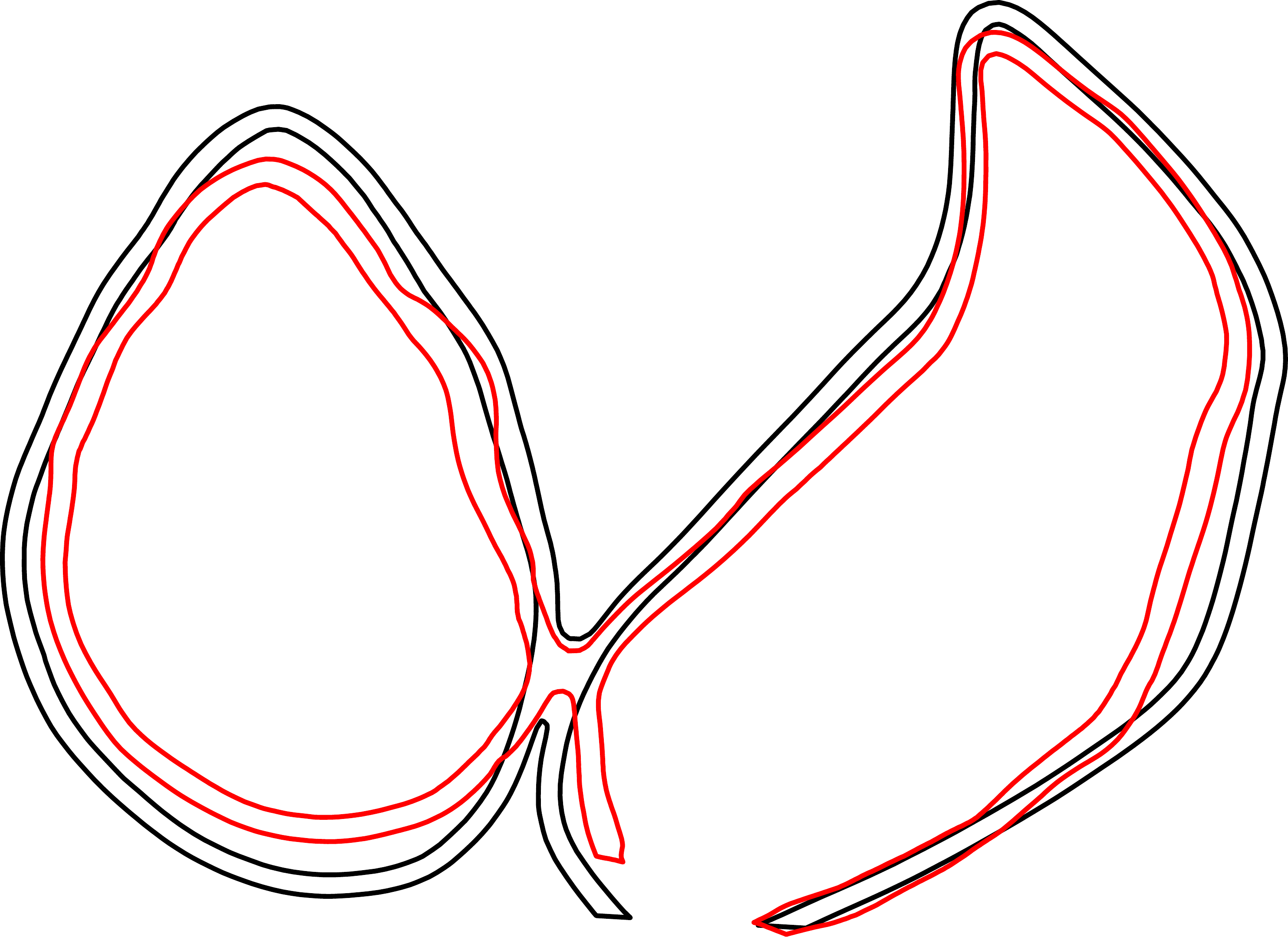}}%
\hfil
\subfloat[Patient 8]{\includegraphics[width=0.15\textwidth]{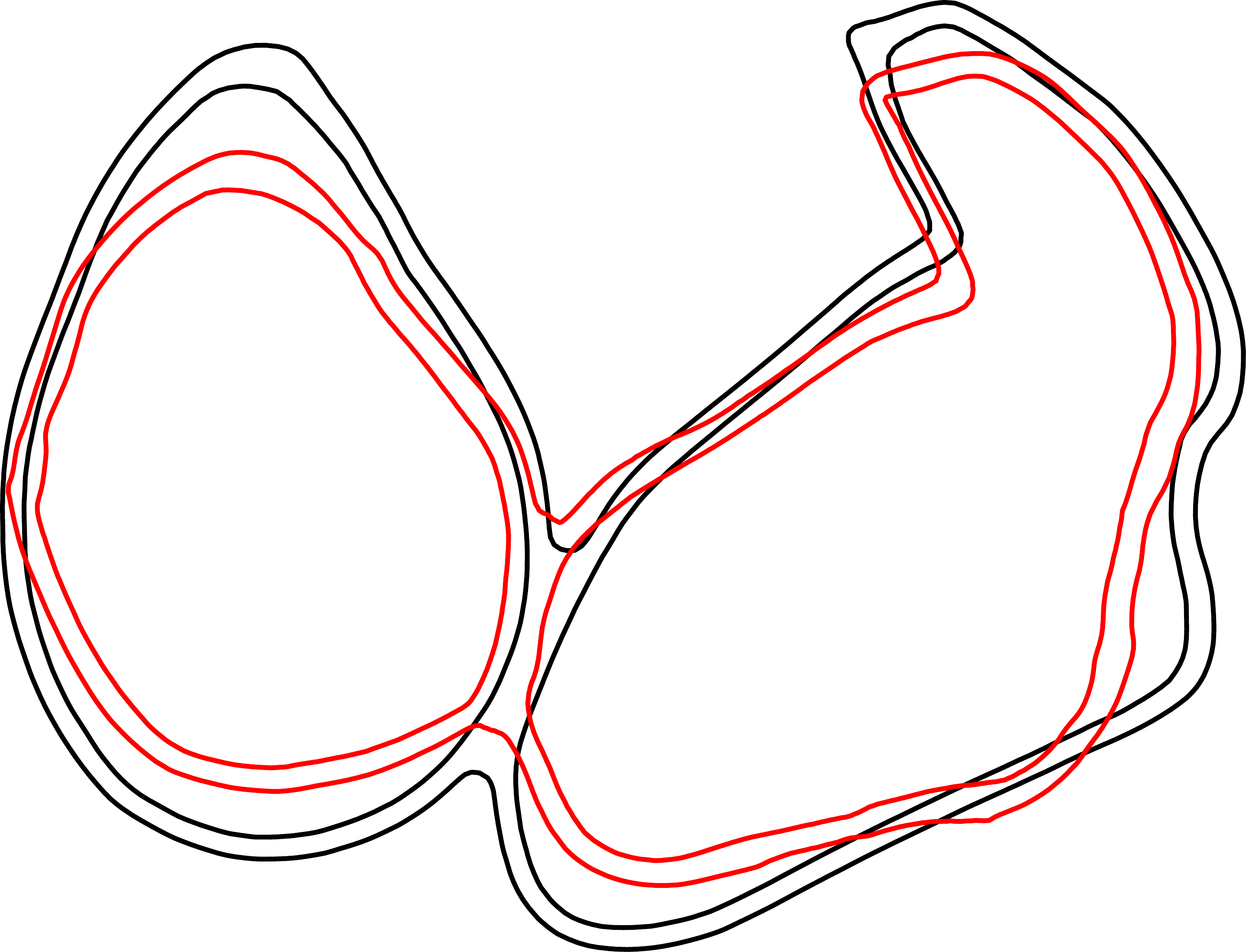}}%
\hfil
\subfloat[Patient 9]{\includegraphics[width=0.15\textwidth]{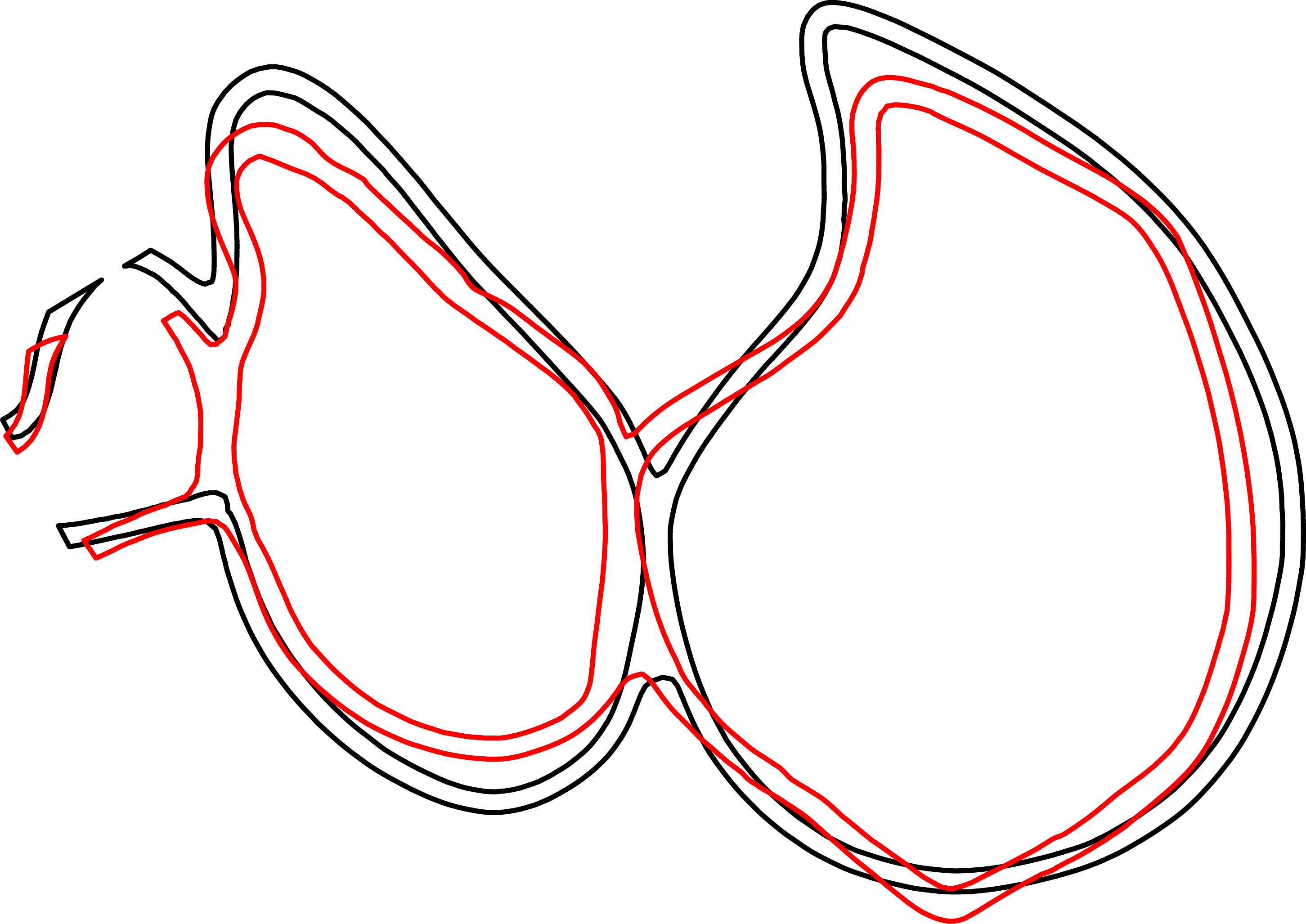}}%
\hfil
\subfloat[Patient 10]{\includegraphics[width=0.15\textwidth]{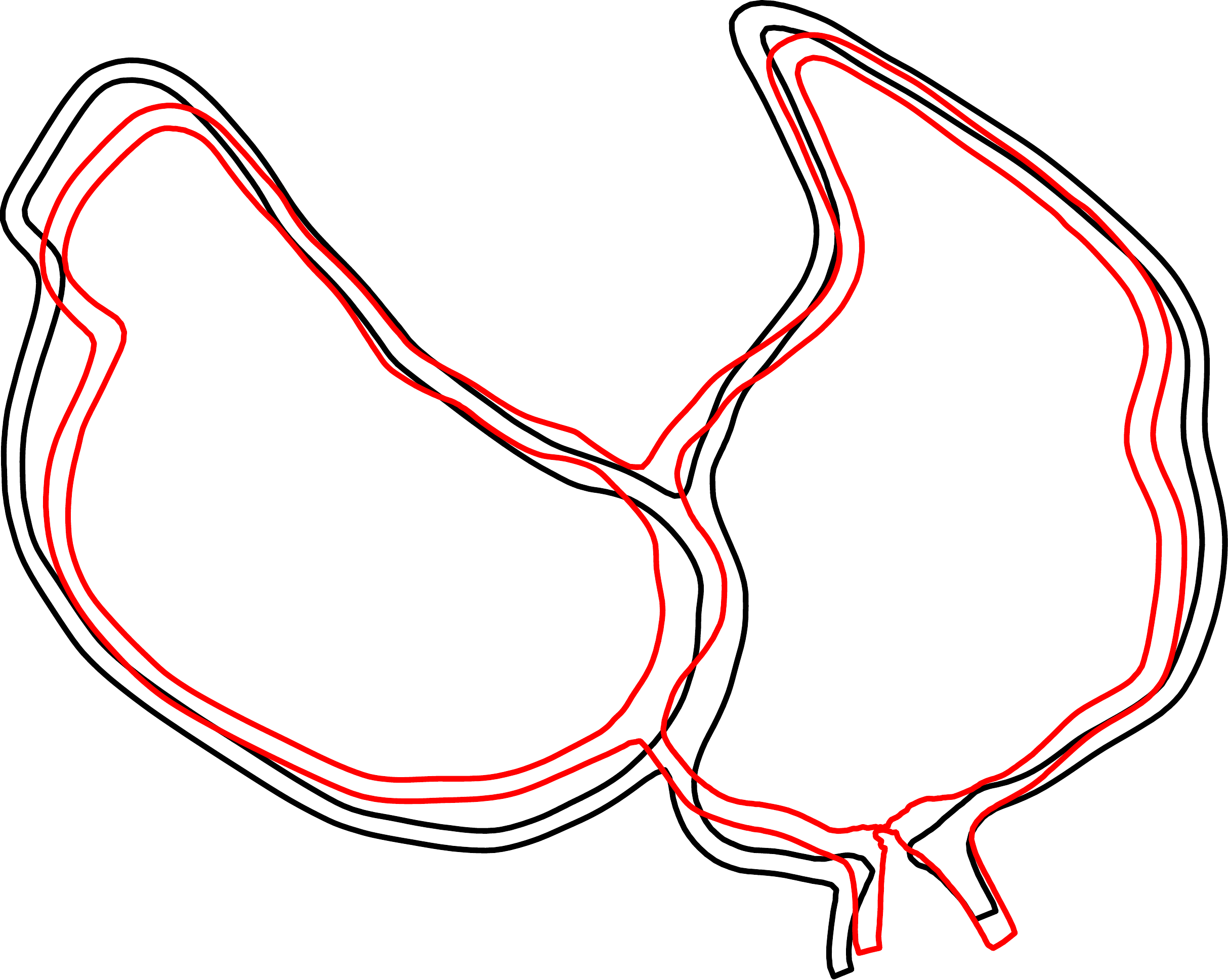}}%
\hfil
\caption{Displacements at maximal contraction of the left ventricle for Patient 2-10 in the 2-chamber-view. The black contour shows the relaxed atria and red contour shows the contracted atria.}
\label{fig_contraction_all}
\end{figure}

Figure \ref{fig_contraction_all} shows the contour lines of the atria of Patient 2-10 at end-systole. All atria perform a physiologically reasonable contraction. The volume and pressure curves until maximal contraction are plotted in Figure \ref{fig_patient_curves}. Despite the differences in volume of the atria, all atria show physiological volume and pressure curves over time. The atrial geometries have been obtained from patients receiving treatment for atrial arrhythmia, thus, some of the geometries are dilated due to sustained atrial arrhythmia and atrial volume is high in some of the patients' atria (see Figure \ref{fig_patient_curves} and Table \ref{tab_values_patient}). Hence, these cases are a good test for our approach. However, our method is able to register the atria and to transfer the fiber orientations although the volume of atlas and patient's atria varies significantly. Additionally, also with such dilated geometry, mechanical contraction can be computed without problems. The ejected volume 
increases slightly with increasing atrial volume, but the ejection fraction decreases in dilated atria due 
to a high initial volume (see Table \ref{tab_values_patient}).

\begin{table*}[htbp!]
\renewcommand{\arraystretch}{1.1}
%if using array.sty, it might be a good idea to tweak the value of
%\extrarowheight as needed to properly center the text within the cells
\caption{Maximal and minimal volume and ejection fraction for Patient 2-10}
\label{tab_values_patient}
\centering
\sisetup{round-mode= places,round-precision = 1}
\begin{tabular}{|l||cc|cc|}
      \hline
 & \multicolumn{2}{c|}{Left}  & \multicolumn{2}{c|}{Right} \\
 &        Max/Min Volume [ml] & EF [\%]   &Max/Min Volume [ml] & EF [\%]    \\
       \hline
  Patient 2 &  \num{114.556698310895} / \num{83.536649356947} & \num{27.4392876676423}  & \num{126.637118441487} / \num{96.658209475075}   & \num{24.1814963949822 }\\
  Patient 3 &  \num{128.04763883753} /  \num{95.701506216093} & \num{25.5692891827454}  & \num{161.743118686833} / \num{119.40086204681}   & \num{26.5429447826017 }\\
  Patient 4 &  \num{157.009621214239} / \num{121.39546503040} & \num{22.9586536344844}  & \num{107.960005404864} / \num{76.763407051905}   & \num{ 29.3700939172109} \\
  Patient 5 &  \num{190.970827833812} / \num{153.11325215410} & \num{20.1183536960575}  & \num{118.767502059093} / \num{85.445548383752}   & \num{28.5108321344938 }\\
  Patient 6 &  \num{145.902620178197} / \num{114.04613013089} & \num{22.1017887345541}  & \num{146.579127704434} / \num{107.08356339605}   & \num{27.4397235027787 }\\
  Patient 7 &  \num{136.207083329857} / \num{103.49447908437} & \num{24.2328496493578}  & \num{150.236727659456} / \num{112.82118863064}   & \num{25.5681164928298 }\\
  Patient 8 &  \num{127.973774311196} / \num{96.511726565111} & \num{24.9034604194291}  & \num{176.428094330608} / \num{136.61722359473}   & \num{23.1800871639543 }\\
  Patient 9 &  \num{96.8237672271759} / \num{69.480117115129} & \num{28.7722180439427}  & \num{147.920611025809} / \num{110.33602711163}   & \num{25.8570572105612 }\\
  Patient 10&  \num{158.223428094065} / \num{126.53332184517} & \num{20.3736621858871}  & \num{136.887713049734} / \num{100.18882307724}   & \num{27.3501772965607 }\\
   \hline
\end{tabular}
\end{table*}

\section{Discussion}
The proposed approach is able to register differently shaped atria to each other and to create physiological reasonable fiber architecture for patient-specific geometries. We show this using ten different human atria with various shapes. Geometric differences in the shape of the atria, as for example the number of pulmonary veins, are handled well. The fibers can be reoriented according to the deformation calculated in the registration process and applied to the patient's atria.

The activation sequence in the patient's atria is similar to the sequence in the atlas atria, thus, it is very important to have an exact fiber definition in the atlas atria.

Using the proposed pipeline to define the fiber orientation on the atria, an exact atlas atria is needed. As the shape of the atria in patients varies very much, an atlas atria should be used which itself is similar to the patient's atria. Although, the method was able to handle a varying number of pulmonary veins, for a better accuracy of the fiber definition at the pulmonary orifices it is suggested to use an atlas with the common number of pulmonary orifices. As for atrial arrhythmia simulations the pulmonary veins play an important role, it could be more convenient to use different atlases with the appropriate number of pulmonary orifices.

Small differences in fiber orientation lead to small differences in the electrical activation pattern. However,  the mechanical contraction showed to be more sensitive to the fiber orientation. 
Individual shapes of the atria, for example additional bulges, cause the registration to deform the atlas atria more to match the individual patient atria, which in turn could lead to strongly varying fiber orientation in the bulges. However, since it is not possible to visualize the fiber orientation in-vivo in the atria, it is not measurable which is the correct fiber definition in these bulges, which appear only in single individuals. Our registration algorithm is able to define fiber orientation everywhere in the atria, i.e. also in bulges. 
However, to improve the performance of the registration and fiber reorientation, different atlas atria could be used for different shaped atria, for example from ex-vivo DTMRI \cite{pashakhanloo_myofiber_2016}. With a similarity measurement between the atlas atria and the patients' atria, for instance the size of the deformation map of the registration, one could find the most similar atlas to be used along with the proposed approach, which would then lead to the most physiological results regarding the fiber orientation and, thus, the activation and contraction. 

\textcolor{black}{The segmentation and the generation of the atrial geometry is restricted by the medical image resolution, which is not enough to capture details of the atrial wall thickness. In the future, a more accurate reconstruction of the atrial wall can be performed during the segmentation process if improved medical imaging is  available. In that case, the proposed method should probably be adapted in order to deal with that level of detail, i.e. build a new atlas atria with the correct wall thickness and segment the wall of patients' atria more detailed.   %Then the more accurate atrial geometry can be straight forward used for the fiber reconstruction without adaption of the methodology.
}

\section{Conclusion}

A local fiber definition is important for patient-specific electrophysiological and mechanical modeling of the atria. %Simulating atrial arrhythmia, appearance, maintenance and termination, without a correct fiber definition the results cannot be utilized for predictions for the patient. 
We presented a method to automatically define the fiber orientation on arbitrarily shaped atria. To do so we use a single pair of atlas atria with a detailed predefined fiber orientation. Using image registration techniques and reorientation methods we are able to map the fibers to different atria, even if the shape of the atria differ significantly. 
The method needs as input only the segmented geometry together with a few landmarks and does not require additional user interaction. Our method is thus very convenient, especially if a study with many individual atria is carried out. 
We compared the result of the fiber mapping with manually defined fibers. The same fiber bundles were visible in defined and mapped atria and the performed electrophysiology and contraction simulation showed similar results for defined and mapped fibers. Using our registration method for fiber mapping to patients' atria, the activation pattern of the patients showed the same characteristics as the atlas. Thus, with a detailed atlas we have the same activation properties also in patient's atria.   
We demonstrated the good performance of our method with ten different human atria. Different shapes of the atria are handled very well during the registration process. 
The electrophysiological, contraction and hemodynamical simulation with atria with mapped fibers showed physiologically reasonable results in terms of activation pattern, spatial contraction, volume and pressure curves, and ejection fraction.

% or


\appendices

% \section{}

\section*{Acknowledgment}

The authors would like to thank the Centers for Radiology of Klinikum rechts der Isar, German Heart Center, Munich, and Kings Collage London, for image data used in this work. Thanks goes also to Martina Weigl and Jonas Niemeijer for their help in segmenting the geometries.

\ifCLASSOPTIONcaptionsoff
  \newpage
\fi

%\paperlayouttriggeratref{8}
%\paperlayouttriggercmd{\enlargethispage{-5in}}


% The paperlayouttran BibTeX style support page is at:
\bibliographystyle{paperlayouttran}
\bibliography{paperlayoutabrv,references}

\end{document}